\documentclass[preprint,superscriptaddress,amsmath]{revtex4-1}
\usepackage{graphicx}
\usepackage{slashed}
\usepackage{bm}
\usepackage{color}
\usepackage[colorlinks,citecolor=blue,urlcolor=blue,linkcolor=blue]{hyperref}

\begin{document}

	\title{Supervised deep learning in high energy phenomenology:\\ a mini review}
	
	\author{Murat Abdughani}
	\email{mulati@itp.ac.cn}
	
	\affiliation{CAS Key Laboratory of Theoretical Physics, Institute of Theoretical Physics, Chinese Academy of Sciences, Beijing 100190, China}
	\affiliation{School of Physics, University of Chinese Academy of Sciences, Beijing 100049, China}
	
	\author{Jie Ren}
	\email{renjie@itp.ac.cn}
	
	\affiliation{CAS Key Laboratory of Theoretical Physics, Institute of Theoretical Physics, Chinese Academy of Sciences, Beijing 100190, China}
	\affiliation{School of Physics, University of Chinese Academy of Sciences, Beijing 100049, China}
	
	\author{Lei Wu}
	\email{leiwu@itp.ac.cn}
	
	\affiliation{Department of Physics and Institute of Theoretical Physics, Nanjing Normal University, Nanjing, 210023, China}
	
	\author{Jin Min Yang}
	\email{jmyang@itp.ac.cn}
	
	\affiliation{CAS Key Laboratory of Theoretical Physics, Institute of Theoretical Physics, Chinese Academy of Sciences, Beijing 100190, China}
	\affiliation{School of Physics, University of Chinese Academy of Sciences, Beijing 100049, China}
	\affiliation{Department of Physics, Tohoku University, Sendai 980-8578, Japan}
	
	\author{Jun Zhao}
	\email{zhaojun@itp.ac.cn}
	
	\affiliation{CAS Key Laboratory of Theoretical Physics, Institute of Theoretical Physics, Chinese Academy of Sciences, Beijing 100190, China}
	\affiliation{School of Physics, University of Chinese Academy of Sciences, Beijing 100049, China}

	\begin{abstract}
		Deep learning, a branch of machine learning, have been recently applied to high energy experimental and phenomenological studies. In this note we give a brief review on those applications using supervised deep learning. We first describe various learning models and then recapitulate their applications to high energy phenomenological studies. Some detailed applications are delineated in details, including the machine learning scan in the analysis of new physics parameter space, the graph neural networks in the search of top-squark production and in the $CP$ measurement of the top-Higgs coupling at the LHC.
	\end{abstract}

	\maketitle
	
	\tableofcontents

	\section{Introduction}
	
	As a typical interdisciplinary field involving mathematical statistics, optimization theory, computer science, algorithm theory and neuroscience, machine learning (ML) mainly studies algorithms and statistic models, which helps improve the performance of computer for given tasks~\cite{webb2003statistical}. For a given task, the ML algorithms can automatically build a mathematical model from training samples and then perform inference directly without any need of manual programming.

	Deep learning~\cite{LeCun:2015}, also known as deep structured learning or hierarchical learning, is part of a broader family of ML methods based on learning data representations, as opposed to task-specific algorithms. Deep learning models are neural networks with many layers, such as the deep fully-connected neural networks, deep belief networks, recurrent neural networks and recursive neural networks. Due to the development of algorithm and computing hardware, deep learning has become a hot topic in both research and application. Actually, deep learning has been applied to many fields (like mail filtering, computer vision, speech recognition, audio recognition, machine translation, bioinformatics, drug design, material design, game design, etc) and demonstrated its learning and inference ability similar or superior to human.

	The traditional ML methods have been applied to the experimental high energy physics (HEP) for more than thirty years~\cite{Bhat:2010zz}. A most successful application is the boosted decision tree (BDT) which helps discover the Higgs boson at the LHC~\cite{Roe:2004na}. Currently, deep learning has been paid great attention not only in experimental HEP, but also in phenomenological studies, e.g.~\cite{Baldi:2014kfa,Baldi:2014pta,Bridges:2010de, Buckley:2011kc, Bornhauser:2013aya, Caron:2016hib, Bertone:2016mdy}. In this note, we give a brief review on these applications. We will first describe various deep neural network models, and then recapitulate their applications to HEP phenomenological studies. Some detailed applications will also be delineated, including the machine learning scan in the analysis of new physics parameter space~\cite{Ren:2017ymm}, the graph neural networks in the search of top-squark production~\cite{Abdughani:2018wrw} and in the CP measurement of the top-Higgs coupling at the LHC~\cite{Ren:2019xhp}. Finally, we give a summary and outlook.

	\section{Machine learning basics}

	The name of machine learning was proposed by Arthur Samuel in 1959~\cite{Samuel:1959}. Then a widely accepted definition for ML algorithm was given by Tom M. Mitchell~\cite{Mitchell:1997}: ``A computer program is said to learn from experience $E$ with respect to some class of tasks $T$ and performance measure $P$ if its performance at tasks in $T$, as measured by $P$, improves with experience $E$". The ML tasks can be classified into supervised learning, semi-supervised learning, unsupervised learning, active learning, reinforcement learning and so on.
	
	\begin{itemize}
		\item Supervised learning algorithms are utilized to build a mathematical model for a data sample with input and target output values~\cite{russell2010artificial}. For a data with partial expected output values, we can use semi-supervised learning to construct a ML model. The typical supervised learning algorithms are for classification and regression. If target output values are discrete categories, it is a classification problem (for example, classification of pictures of different kinds of cats). If target output values are continuous, it is a regression problem (for example, prediction of house prices according to their features).

		\item Unsupervised learning algorithms are utilized to construct a mathematical model for a data sample with only inputs but no target output values~\cite{hinton1999unsupervised}. Unsupervised learning algorithms are usually used to find out the hidden structures in a data sample, e.g., to cluster the data into groups according to some given criteria).
		
		\item Active learning algorithms can reduce the total cost of acquiring data sample, especially when the cost of labeling the target output values of a given input data is very high. Active learning can preferentially pick out those data samples which need to be labeled and thus lowers the total cost of acquiring data sample.
		
		\item Reinforcement learning~\cite{wiering2012reinforcement} algorithms are mainly used to deal with dynamical environment problems. In this kind of problems, data mainly comes from observing environment state and the reward given by environment after some actions performed. For example, in various chess games the environment is the current state of chessboard, i.e., the positions of chess pieces. Moving a chess piece is an action which will lead to a new chessboard state. The final score, win or lose, is the reward given by environment after performing a series of actions.
	\end{itemize}

	ML has a close relation with optimization theory. They have something in common. Most ML problems use loss function to measure the difference between output values and target values, and thus the ML problems can be written as problems of minimizing the loss function. The difference of ML from optimization theory is that, although optimization theory can minimize the loss of ML model over the training data set, ML pay more attention to the loss of ML model over the data set with unknown target output~\cite{Roux:2012}. In the ML field, we use generalization performance to measure the performance of ML models over new data or environment not participating in the learning. Therefore, optimizing the generalization performance is the main goal of ML.

	ML also has a close relation with statistics. In ML, from mathematical models to theoretical tools, there are many correspondences in statistics. Some scholars think that ML emphasizes studying algorithm models while statistics emphasizes data models. The ML field which mainly applies statistical methods is usually called statistical learning~\cite{webb2003statistical}.

	\subsection{Supervised learning}
	
	In general, in order to solve a given supervised learning problem, we need the following steps:
	
	\begin{enumerate}
		\item Determine the type of training sample.	
		For example, to train a ML model for handwriting character recognition, the samples we collected are pictures of various handwritten characters, handwritten words, or whole lines and paragraphs of handwritten words.

		\item Collect data samples as the training set.	
		The data samples in the training set should be representative and can reflect the data distribution in the real application environment. Meanwhile, for each data sample we need to label its corresponding target output value (the expected value output from ML model). Such data labeling can be done by human experts or by various measurement methods.
		
		\item Design the features of data sample as input of ML model.	
		The accuracy of a ML model depends heavily on how data samples are represented. Usually, a data sample is represented as a high-dimensional feature vector whose each component describes a specific aspect of the data sample. In order to improve the accuracy of ML model, a feature vector must contain enough information on the data sample. However, due to the curse of dimensionality, the number of features should not be too larger.
		
		\item Choose a ML model and its corresponding learning algorithm.	
		For example, the commonly used supervised ML models include the generalized linear regression, logistic regression, decision trees, support vector machine and neural networks.
		
		\item Train the ML model with the collected training samples.	
		Generally, ML models and supervised learning algorithms have some hyperparameters to control the complexity of the model and the details of learning. These parameters are usually adjusted according to the generalization performance of the ML model on a separate validation set, which is called cross validation.
		
		\item Assess the accuracy of ML model.	
		After the adjustment of hyperparameters and the optimizing the learnable parameters in the ML model, the accuracy of the final ML model needs to be further evaluated on a separate test set.
	\end{enumerate}

	We now give the working principle of supervised learning algorithm without loss of generality. Given a training set containing $N$ samples, and each sample has an input feature vector $\bm{x}_i \in X$ and a target output value $\bm{y}_i \in Y$. Here, we assume that the samples in the training set are independently and identically distributed, and they are also independently and identically distributed with the data samples in the real application environment. The task of supervised learning is to find a function mapping $g: X \to Y$ between the input feature space $X$ and the target output space $Y$. We call the function space formed by all the candidate functions $g$ as the hypothesis space.

	In order to evaluate the goodness of any function $g$ fitted to the sample, we need to define a loss function $L$. When the function $g$ gives the output value $\hat{\bm{y}}_i$ for training sample $(\bm{x}_i,\bm{y}_i)$, the function $g$ has a prediction error $L(\bm{y}_i, \hat{\bm{y}}_i)$ on this sample. For the entire training set, the empirical risk of function $g$ is defined as
	\begin{equation}
		R(g) = \frac{1}{N} \sum_i L(\bm{y}_i, g(\bm{x}_i)) .
	\end{equation}

	There are two ways to select the function $g$. One is to minimize the empirical risk, and the other is to minimize the structural risk. With empirical risk minimization, we can turn the supervised learning problem into an optimization problem that minimizes $R(g)$. In this case, if the function $g$ is a conditional probability $P(\bm{y}|\bm{x})$ and the loss function is chosen to be the negative logarithmic likelihood function $L(\bm{y}_i, \hat{\bm{y}}_i) = -\log P(\bm{y}|\bm{x})$, minimizing the empirical risk is equivalent to maximizing the likelihood function. The downside of this approach is that if there are only small number of training samples and the hypothesis space is too large, the function $g$ will over-fit the data seriously, resulting in a poor generalization performance.

	By introducing penalty terms to regularize the model, the structural risk minimization can suppress the over-fitting problem of empirical risk minimization. The penalty terms constrain the complexity of the function. Penalty terms can take a variety of forms. Denoting the learnable parameters in function $g$ as $\{\beta_j \}$, the Euclidean norm ($L_2$ norm) for these parameters $\lVert \bm{\beta} \rVert_2 = \sum_j \beta_j^2$ can be used as the penalty term, which is also the most commonly used penalty term. Other norm forms of learnable parameters, e.g., $L_1$ norm $\lVert \bm{\beta} \rVert_1 = \sum_j |\beta_j|$ and $L_0$ norm (the number of non-zero $\beta_j$) are also the commonly used penalty terms. For the sake of discussion, we denote all the penalty terms as $C(g)$. In this way, we can transform the supervised learning problem into an optimization problem that minimizes structural risks
	\begin{equation}
		J(g) = R(g) + \lambda C(g) .
	\end{equation}
	Here the constant parameter $\lambda$ controls the prediction bias and variance of the function $g$. When $\lambda = 0$, $J(g)$ is the empirical risk $R(g)$ and the optimized function $g$ has a smaller prediction bias and a larger prediction variance. When $\lambda$ is too large, the optimized function $g$ has a large prediction bias and a smaller prediction variance. In general, we use cross validation to select an appropriate $\lambda$ to maximize the generalization accuracy of the function $g$.

	In fact, choosing a ML model is to select a function with tunable parameters ${\beta_i}$, which constitutes the hypothetical space $G$. The purpose of supervised learning algorithm is to minimize the generalization risk of the ML model by adjusting the learning parameters in the model. Usually, the risk function is also called the loss function.

	So far various ML models have been developed. However, there is no single ML model that can cope all the types of data and learning tasks. Each ML model has its pros and cons. Therefore, to select a ML model, we need to consider at least the following factors:
	\begin{itemize}
		\item The trade off between prediction bias and variance.	
		The prediction error of ML model can be decomposed into bias and variance. If a ML model has a small prediction error in the validation set, it means that the ML model is flexible enough to learn data well. But if a ML model is too flexible, it will learn the training set too well and perform poorly on the validation set, resulting in a too large prediction variance.
		
		\item The complexity of the data and the number of training samples.	
		If the hidden pattern of the data is simple, a simple ML model can describe the data well by learning a small number of training samples. Otherwise, if the data distribution function is quite complex, a large training set, many features of each training sample and a much flexible ML model are needed.
		
		\item Number of input features.		
		If the dimension of input feature vector is very high but the actual data distribution is simple, the extra data dimension will lead to a large prediction variance of ML model, and thus increasing the difficulty of training ML model. In order to avoid such problems, we usually preprocess the data, such as using dimension reduction technique to remove irrelevant features to reduce the dimension of the input space, so as to improve the prediction accuracy of the ML model.
		
		\item Noise in the target output of samples.	
		If the target output values are too noisy, the ML algorithm will try to learn the noise in the training sample, thus making the generalization performance of the ML model in the validation set worse. This phenomenon is called over-fitting. Even if there is no noise in the target output of samples, when the flexibility of the ML model itself exceeds the complexity of the data to be learned, it will also cause the over-fitting problem. In the actual ML model training, we usually use techniques such as early stopping to detect and reduce the influence of noise in the training samples on the generalization performance of ML.
	\end{itemize}

	\subsection{Decision trees}
	
	Decision trees are a kind of non-parametric statistical ML models with tree structure, which can be used to deal with both classification and regression problems~\cite{shalev2014understanding}. If the value of the output is discrete, the decision tree is called a classification tree. Otherwise, if the output is continuous, the decision tree is called a regression tree.
	
	\begin{figure}[t!]
		\centering
		\includegraphics[width=14cm]{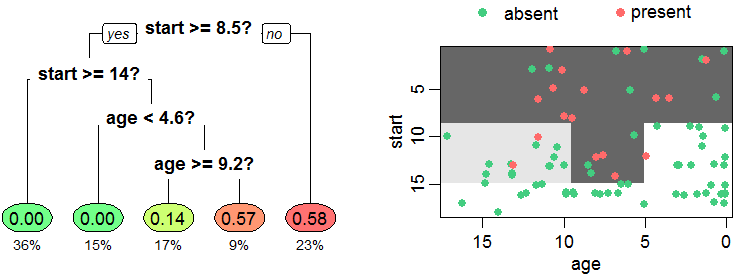}
	        \caption{A decision tree that estimates the probability of a patient developing a hunchback after surgery based on the patient's age and the location of the spine where the surgery was performed. The figure shows two different representations of the decision tree. The colored leaf nodes in the left figure give the probability of postoperative hump and the proportion of samples reaching each leaf node. The right figure shows the age of the patient and the position of the spine on which the surgery was performed, where different regions correspond to different leaf nodes. The locations of the sample points are also shown as points.  This picture is taken from https://en.wikipedia.org/wiki/Decision-tree-learning.}
		\label{Chap2/dt-space}
	\end{figure}

	Take a decision tree shown in the left panel of Fig.\ref{Chap2/dt-space} as an example. Each non-leaf node in the tree corresponds to an input variable. It has two (or more) edges that connect the child nodes, and each edge corresponds to a range of values for the input variable. Each leaf node labels an output value. When a sample starts from the root of the tree and reaches a leaf node by following the branches determined by the non-leaf nodes, then the predicted output value of the sample is the value attached to the leaf node.
	
	Decision tree is a learnable model. As shown in the right panel of Fig.\ref{Chap2/dt-space}, the purpose of decision tree algorithms are to learn from training samples some classification rules, which divides the input feature space into many regions.	These classification rules are the non-leaf nodes and edges in the decision tree, and the regions are correspond to the leaf nodes.
	
	The advantages of a decision tree are~\cite{james2014introduction}:
	(i) They are white box models, easy to understand, explain and visualize;
	(ii) Few data preprocessing are needed;
	(iii) Low computational cost to make an inference;
	(iv) Can process both category data and numerical data;
	(v) Can handle multiple output variables.
	(vi) The reliability of decision trees can be validated by statistical test;
	(vii) Even if the real data and training samples are not completely consistent, the decision trees usually still have good generalization performance.
	
	The disadvantages of a decision tree are:
	(i) The learning algorithms may construct over-complex decision trees, which leads to poor generalization performance;
	(ii) The decision trees can be unstable;
	(iii) The problem of learning optimal decision tree is NP-complete~\cite{HYAFIL197615};
	(iv) There are some logical relationships that are not easily represented by decision trees;
	(v) If the number of training samples in each output category is unbalanced, the decision trees constructed by the learning algorithms are biased.
	
	In order to solve the shortcomings of a single decision tree, we usually combine decision trees in the ensemble learning framework. Taking the classification problem as example, the purpose of ensemble learning is to improve the overall generalization performance and prediction stability by combining the output values from multiple weak classifiers. There are usually two kinds of ensemble learning methods. The first one is to take the average. This type of methods build multiple weak classifiers $\{C_k\}$ independently and then simply average their outputs as the overall output value $\bm{y} = \frac{1}{N} \sum_k C_k(\bm{x})$, where $\bm{x}$ and $\bm{y}$ are input and output of decision trees, respectively. Because taking the average can reduce the variance of the output value, the overall prediction stability of multiple classifiers is better than a single classifier. Such methods include bagging and random forest, etc.
	
	The second kind of methods are boosting. They build each weak classifier sequentially. Each new weak classifier $C_k$ is constructed to reduce the prediction bias of the combined output values given by all the previously constructed weak classifiers $C_1, \dots, C_{k-1}$. AdaBoost~\cite{Freund1997} and gradient tree boosting are two commonly used models. The built decision trees are generally called boosted decision trees (BDT), which is one of the most widely used ML models in HEP experiments.
	
	\section{Deep neural networks}
	
	Deep learning is a branch of neural network~\cite{Bengio:2009, LeCun:2015}. In deep neural networks, each neural network layer learns a transformation of the data output from the previous layer. The representation of data becomes more abstract as the depth of the network increases, making it more efficient to learn the complex correlations between features. Take human face recognition as an example. the original input image is usually represented as a matrix composed of pixel luminosity. The first neural network layer will focus only on learning the correlations between local pixels in the image, which extracts the edges from the image. The second layer is used to learn how to combine and encode the pattern of edges, and to identify the noses and eyes patterns. The last layer will be used to classify whether the image contains a human face. Therefore, the key point of deep learning is that it puts the learning and optimization of different levels of data abstraction into many network layers, and different layers produce abstract data representations in different levels. In this way, deep learning can realize end-to-end learning, from the raw input data to the target output, without the manual design of input features. In 2017 the depth of neural network model can reach thousands of layers, and the number of neurons and connections can reach millions. Although that number is many orders smaller than the human brain, neural networks can outperform humans at certain tasks, such as the game of go.

	Many kinds of deep neural network structures have been developed. The commonly used deep neural networks include fully-connected neural network, recurrent neural networks (RNN), convolutional neural networks (CNN), etc. Each deep learning model can achieved great success in its applicable field.

	While deep neural networks have been hugely successful, they also have their own problems. The two biggest problems are the serious over-fitting and heavy computational consumption. (i) Since a deep neural network has many layers, it will also learn from the noise in the training set, which leads to serious over-fitting problem. Various regularization methods, such as neuronal pruning~\cite{Ivakhnenko1971}, weight decay ($L_2$ regularization) and sparsification ($L_1$ regularization), are used to prevent over-fitting during training~\cite{Bengio2013}. In addition, Dropout regularization could prevent over-fitting by randomly restraining some neurons in hidden layers during training \cite{Dahl2013}. There are many other methods to reduce over-fitting, such as expanding the training set by clipping, rotating and flipping images. (ii) Deep neural network has many hyperparameters that need to be optimized, such as the number of layers in the neural network, the number of neurons in each layer, the learning rate of training, and the initial value of learnable parameters. Because of the limitation of time and computing resources, it is not practical to find the optimal values in the whole space of the hyperparameters. Various techniques are used to speed up the training of deep neural networks. For example, the gradient of loss function can be calculated with small batches of training samples instead of the whole training set~\cite{Hinton2012}. In addition, because the deep neural networks involve a large number of matrix and vector calculations, utilizing large-scale multi-core computing architectures, such as NVIDIA graphics processing unit (GPU) and Intel Xeon Phi, can significantly improve the training speed of deep neural networks~\cite{You:2017, Viebke2017}.

	\subsection{Fully connected neural network}

	Artificial neural network is a computing system inspired by biological neural network, which belongs to connectionism model. Similar to a biological neural network, an artificial neural network consists of a series of artificial neurons that communicate with each other through connections. The artificial neuron processes the received data and passes the processing results to the subsequently connected artificial neurons. Artificial neurons can be stateless or have a memory. Typically, artificial neurons are organized in layers, with different layers of neural networks, which applys different transformations to the data. Data are input from the first layer, transformed by each layer, and output from the last layer. For the convenience of description, the name ``neural network" refers to artificial neural network in the following.

\begin{figure}[t!]
		\centering
		\includegraphics[width=12cm]{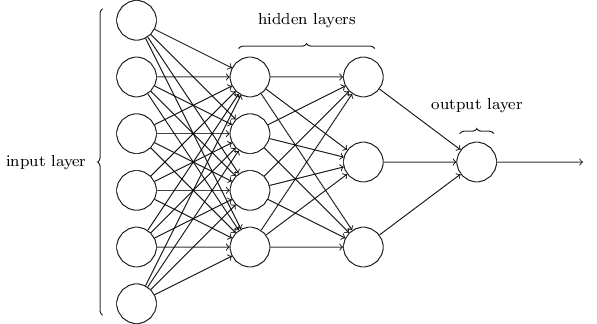}
		\caption{Diagram of a fully connected feed-forward neural network with one hidden layer. This picture is taken from https://computersciencewiki.org/index.php/Multi-layer\_perceptron\_(MLP).}
		\label{Chap2/mlp}
	\end{figure}

	Fully connected neural network (FCNN) is the simplest neural network model, which is a ML model that can learn function mapping of any vector to vector through training. As shown in Fig.~\ref{Chap2/mlp}, neurons in FCNN have a layered structure. The first layer is the input layer, and the last layer is the output layer. Between the input and output layers are hidden layers. The input layer contains $N_i$ neurons, each corresponding to an input feature of the sample. Each neuron in the hidden layer takes the output of neurons in the previous layer as a weighted linear sum and applies nonlinear transformation as the output, where the nonlinear transformation function is called the activation function of a neuron. Usually a layer of neurons have the same activation function.

	The number of hidden layers in a FCNN, the number of neurons in each hidden layer, and the activation function of each hidden layer are all chosen depending on the specific learning tasks. They are selected through cross-validation or other methods, according to the complexity of the problem to be learned and the size of the training set, etc. The commonly used activation functions include sigmoid function, tanh function, relu function, leaky relu function, etc. In addition, we usually use the term ``multilayer perceptron" to refer to a FCNN with only one hidden layer~\cite{Hastie2009}.

	FCNN has several advantages, e.g., able to learn nonlinear mapping and able to perform on-line learning (by acquiring more training samples, the learned nonlinear mapping relationship is gradually improved to improve the prediction accuracy). FCNN also has some drawbacks. First, the loss function of a FCNN is usually nonconvex and has many local minima. Therefore, different initial values of learnable parameters in the neural network will lead to completely different parameter values, further affecting the generalization prediction accuracy. Second, FCNN has many hyperparameters that need fine tuned, such as the number of hidden layers, the number of neurons in each hidden layer, and the number of training iterations. Third, FCNN is very sensitive to the scale of input features. It is usually necessary to first normalize the input features of the samples.

	\subsection{Convolutional neural network}

	Convolutional neural network (CNN) is mainly applied to the analysis of image data. It is a deformation of multilayer perceptron, mainly used to reduce the image preprocessing work. Such networks are characterized by weight sharing and image translation invariance~\cite{Zhang1988, Zhang1990}. Inspired by the biological information processing~\cite{Matsugu2003}, the connection pattern of neurons in CNN imitates the connection of biological neurons in biological visual cortex. A neuron in a biological visual cortex responds only to visual stimuli in a certain region known as the reception field of the neuron. The reception fields of adjacent neurons overlap, and the combine reception field of all neurons covers the entire field of vision.

	There is almost no need to manually preprocess the image. The CNN can learn how to filter the image step by step. In this way, the learning of CNN is not limited by human prior knowledge, and the inference performance is greatly improved. CNN has been widely used in image and visual recognition, recommendation system~\cite{Oord2013}, image classification, medical image analysis and natural language processing~\cite{Collobert2008}. A very successful application is the AlphaGo~\cite{Silver2017}, which takes the state of the go board as an image and uses CNN to extract its features. Together with the use of Monte Carlo tree search, reinforcement learning and other techniques, AlphaGo beats the top human players for the first time.

	\begin{figure}[t!]
		\centering
		\includegraphics[width=14cm]{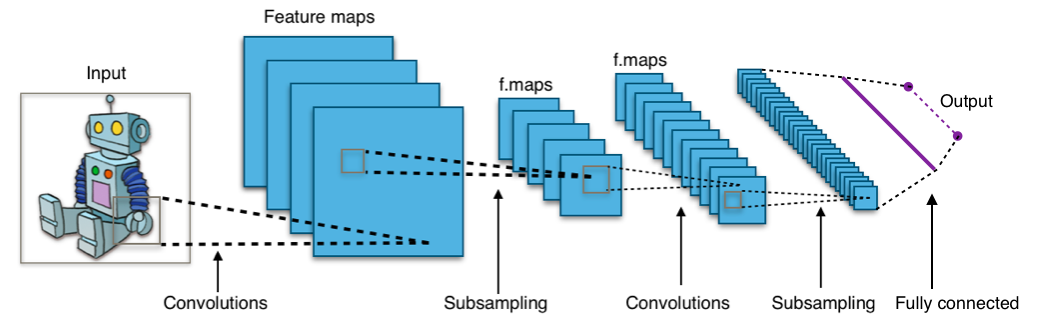}
		\caption{Schematic diagram of a typical structure of convolutional neural network. This piture is taken from https://en.wikipedia.org/wiki/Convolutional\_neural\_network.}
		\label{Chap2/cnn}
	\end{figure}

	Fig.~\ref{Chap2/cnn} shows a typical CNN structure. CNN is different from the fully connected NN in that the hidden layers of the CNN usually include convolutional layer, ReLU and other nonlinear activation functions, pooling layer, full-connected layer and normalization layer, etc.
	\begin{itemize}
		\item The convolutional layer is used to apply a convolution operation (actually a cross correlation operation) to the input image and pass the processing results to the next layer of the NN. This convolution operation simulates the response of each neuron to visual stimuli in its reception field. The size of the convolution kernel determines the reception field of the neuron. If the fully connected NN is used, a neuron needs to be connected to every pixel of the input image, so a lot of neural connections are needed to process the input image, thus forming a large number of learnable parameters. However, when using the convolution layer, only the parameters of the convolution kernel need to be learned, and the size of the convolution kernel is usually small, which greatly reduces the number of parameters to be learned in the NN. In addition, it is worth noting that the convolution operation is translation invariant for image processing, and different neurons in the same convolutional layer share the same convolution kernel. Therefore, it is an implementation of local connection and weight sharing in the convolutional layer.
		
		\item The pooling layer is used to combine the outputs of multiple neurons~\cite{Ciresan:2011, Krizhevsky:2017}. For instance, maximum pooling~\cite{Mittal2018} computing takes the maximum output of a group of neurons as the total output, while average pooling~\cite{Ciresan2012} computing takes the average output of a group of neurons as the output.
		
		\item The fully connected layer is the same as the NN layer in the fully connected NN.
	\end{itemize}

	Although having fewer learnable parameters than the fully connected NN, CNN has a great performance advantage in the field of image data processing, in order to avoid the problems caused by over-fitting. But we still need a large number of training samples to train the CNN~\cite{Maitra2015}.

	\subsection{Recurrent neural network}

	In the feed-forward NN, we assume that all input data are independent. For sequential data, this assumption is unreasonable. For example, in a natural language sentence, there may be a strong connection between two words, even they are far apart. For this kind of problem, we can treat the input data as a sequence to be processed by the recurrrent neural network (RNN). Unlike the feed-forward NN, the neurons in the RNN have memories stored in their internal state vectors. The current data processing depends on the results of the previous calculation, which enables the RNN to process the information in the sequence data very well.

	\begin{figure}[t!]
		\centering
		\includegraphics[width=14cm]{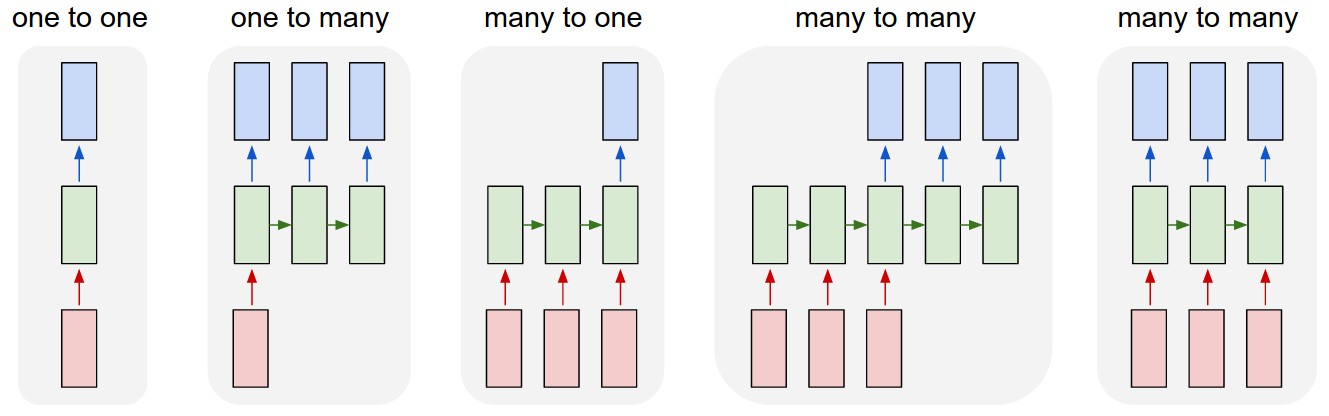}
		\caption{The types of sequential tasks that recurrent neural networks can handle. Each rectangle represents a vector, and the arrow represents applying transformation. Input vectors, output vectors, and network state vectors are represented in red, blue, and green, respectively. This piture is taken from http://karpathy.github.io/2015/05/21/rnn-effectiveness.}
		\label{Chap2/rnn-task}
	\end{figure}

	As shown in Fig.~\ref{Chap2/rnn-task}, learning tasks can be classified into several categories according to the relationship between input data and output data.
	\begin{enumerate}
		\item Single data input to single data output, which can be solved by using feed-forward NN;
		\item Single data input to sequence data output, for example, given an image and generate the text description of this image;
		\item Sequential data input to a single data output, such as determining whether the content of a given text paragraph is a positive comment;
		\item Sequence data input to sequence data output, for example, a sentence in English language is translated into a sentence in Chinese language;
		\item Synchronous sequence data input to sequence data output, such as classification of each frame in video.
	\end{enumerate}

	\begin{figure}[t!]
		\centering
		\includegraphics[width=10cm]{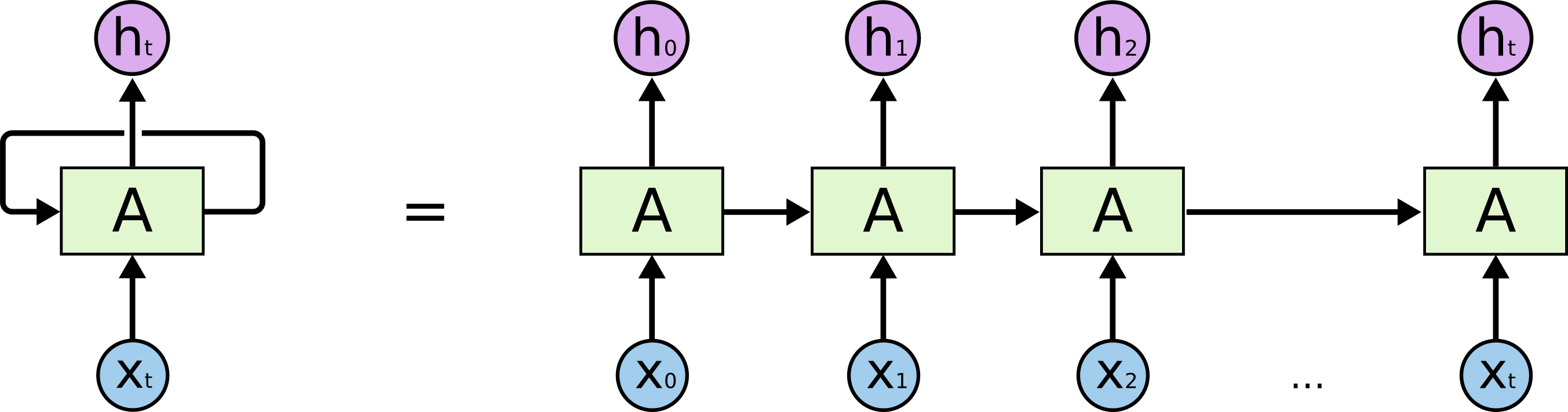}
		\caption{A schematic diagram of the structure of the recurrent neural network and the complete neural network expanded along the iteration time. This piture is taken from http://colah.github.io/posts/2015-08-Understanding-LSTMs.}
		\label{Chap2/rnn-unfold}
	\end{figure}

	The left side of Fig.~\ref{Chap2/rnn-unfold} illustrates the structure of a typical RNN. For the convenience of analysis, the RNN is usually expanded into a complete NN along the iteration time. As shown on the right side of Fig.~\ref{Chap2/rnn-unfold}, its connection pattern forms a directed graph along time. When the sequence length of the input data is $T$, the time-expanded NN has $T$ layers, where the input data of the neuron in $t$-th layer is $\bm{x}_t$ and the output data is $\bm{h}_t$. In each iteration, the original RNN neuron performs the following nonlinear transformation
	\begin{equation}
		\bm{h}_t = \tanh( W \bm{x}_t + U \bm{h}_{t-1} + \bm{b}) .
	\end{equation}
	However, the use of RNN neurons results in too much focus on the most recent data input, so it cannot well describe the long-distance correlation in the sequence, that is the correlation between the data with a longer distance. Therefore, the long short-term memory (LSTM) \cite{Hochreiter:1997} and gated recurrent unit (GRU) \cite{ChoMGBSB14} are developed.

	In theory, RNN can process sequence data of any length. Due to the great advantages of RNN in processing these sequence problems, its various deformations have been widely applied in many fields such as online handwritten character recognition~\cite{Graves2009}, speech recognition~\cite{Sak2014, Li2014}, natural language understanding, image description generation and so on.

	\subsection{Recursive neural network}

	Recursive neural network (RecNN) is a kind of deep learning model applied to data of tree structure. The RecNN was first used for structural decentralization~\cite{Christoph:1996}, and then it was further developed in the 1990s~\cite{Sperduti:1997, Frasconi:1998}.

	\begin{figure}[t!]
		\centering
		\includegraphics[width=8cm]{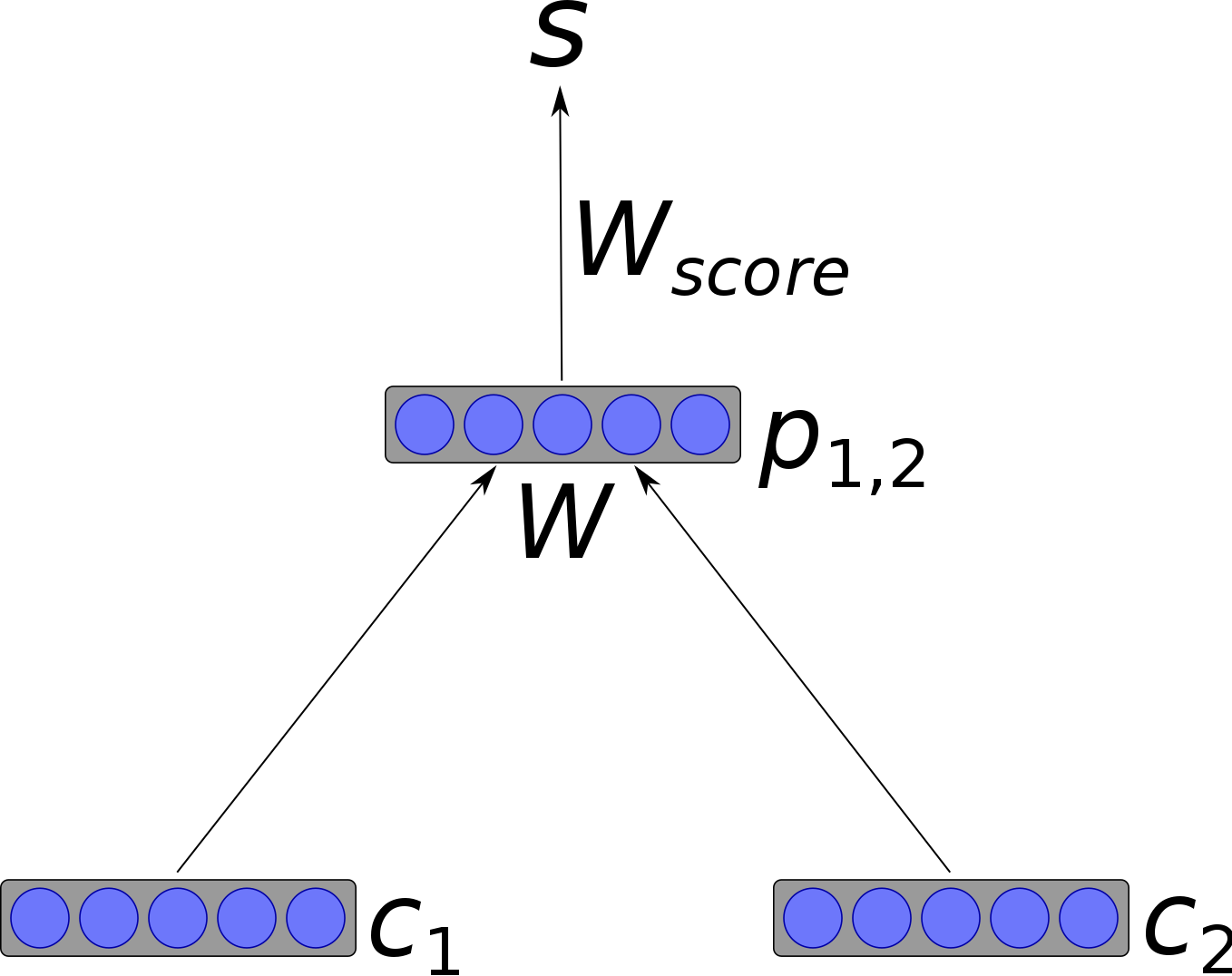}
		\caption{Typical structure of binary recursive neural network.
			The picture is taken from https://en.wikipedia.org/wiki/Recursive\_neural\_network.}
		\label{Chap2/ml-recnn}
	\end{figure}

	Fig.~\ref{Chap2/ml-recnn} gives a typical binary RecNN. This structure performs nonlinear transformation of the child-node features $\bm{c}_i$ and $\bm{c}_j$ to father-node data $\bm{p}_{ij}$, for example,
	\begin{equation}
		\bm{p}_{ij} = \tanh(W [\bm{c}_i; \bm{c}_j] + \bm{b}) ,
	\end{equation}
	where $[\cdots]$ represents vector concatenation, and $W$ and $\bm{b}$ are learnable parameters. Starting from the leaf nodes, such a transformation is applied to every node in the tree until reaching the root node. Then the features of the root node is a compact vector representation of the whole tree structure. Similar variants of RecNN have been widely used in the fields of natural scene understanding and semantic analysis of natural language.

	\subsection{Graph neural network}

	Graph neural network (GNN) is a kind of deep learning model applied to data of graph structure. On the basis of the GNN model, the deep tensor network~\cite{Schuett2017}, message passing neural network (MPNN)~\cite{Gilmer2017}, locally connected neural network~\cite{Bruna13}, graph convolutional neural network~\cite{NIPS2015_5954, Defferrard16, Kipf16} have been also developed.

	For the sake of simplicity, we consider only the MPNN for undirected graphs. Denote the feature vector of each node as $\bm{x}_v$ and the feature vector of each edge as $\ bm{e}_ {vw}$ in an undirected graph $G$. In the embedding layer, each node generates its hidden states $\bm{h}_v^0$ according to its feature vector. Typically, A certain number of message passing iterations are performed on the input graph. In the $t$-th message passing, the hidden states $\bm{h}_v^t$ of each node in the graph are updated according to the following equations,
	\begin{eqnarray}
		\bm{m}_v^{t+1} &=& \sum_{w \in G} M_t(\bm{h}_v^t, \bm{h}_w^t, \bm{e}_{vw}) \\
		\bm{h}_v^{t+1} &=& U_t(\bm{h}_v^t, \bm{m}_v^{t+1}) 
	\end{eqnarray}
	where $M_t$ is a learnable message function and $U_t$is a learnable node status update function. If the messags passing runs totally $T$ iterations, then the hidden states of each node in the graph are $\bm{h}_v^T$. Finally, data is read from all (or part) of nodes in the graph
	\begin{equation}
		\bm{y} = R({\bm{h}_v^T} | v \in G)
	\end{equation}
	as the output value of the neural network to the input graph. Here $R$ is a learnable readout function.

	\section{Machine learning in high energy phenomenology}

	\subsection{A brief overview}

	Machine learning is a powerful tool for the study of high energy physics, because it can discover the hidden patterns in a large amount of complex data and apply them to new data or physical models~\cite{Roe:2004na}. One of the big successful applications is the use of the Boosted Decision Trees (BDT) in the discovery of the Higgs boson in the LHC experiments~\cite{Bhat:2010zz}. In recent years, more and more advanced machine learning techniques, such as Deep Neural Network (DNN) and Message Passing Neutral Network (MPNN), have been applied to the research of new physics experiments and phenomenology, e.g.~\cite{Baldi:2014kfa,Baldi:2014pta,Bridges:2010de, Buckley:2011kc, Bornhauser:2013aya, Caron:2016hib, Bertone:2016mdy}. We review such applications in the following.

	\subsubsection{Parameter scan}

	The ML model has been used to learn and approximate the mapping between the parameter space of the new physics model and the experimental physical observables, so as to efficiently constrain the parameter space of the new physics model.

	\begin{itemize}
		\item {\bf Learning new mass spectrum using multilayer perceptron.}
		Under the sensitivity of ATLAS experiment, it needs a very large amount of computing power to deduce the survived regions of the parameter space of CMSSM using Bayesian posterior probability and likelihood function ratio test. In order to reduce the amount of computation, the article~\cite{Bridges:2010de} uses a multilayer perceptron as a regressor to learn the mapping from the CMSSM model parameter $\bm{\theta}$ to the weak-scale supersymmetric particle masses $\bm{m}$. The output of physical package SoftSusy~\cite{Allanach:2001kg} was used as the target output value of the neural network. About 4000 sample points in the parameter space were collected as the training set to train the regressor. Given a set of CMSSM parameters, this multilayer perceptron model was used to rapidly predict the corresponding supersymmetric particle mass spectrum. It was found that this method is much faster than the traditional calculation method by about $10^4$ times.

		\item {\bf Using multilayer perceptron and support-vector machine to learn the number of new physics events at the collider.}
		It can be very time consuming to generate collider event samples at the LHC using Monte Carlo simulation. A fast detector simulation usually takes a few minutes, while a full detector simulation based on the experimental group GEANT4 would take several days. Therefore, to study the supersymmetric model, the ATLAS and CMS experimental groups limited the sample points of model parameters to a small number of grid points on certain planes, and then perform parallel full detector simulations with these parameters. If a ML model can be constructed to learn the mapping between the parameter points of the supersymmetric model and the output results of the collider simulation, we can generalize this to the entire parameter space of the supersymmetric model and thus the number of signals generated by the specific model at the LHC collider can be estimated in a very short time. In this way, the phenomenological study can be accelerated. In~\cite{Buckley:2011kc} the CMSSM model with four parameters was chosen to be investigated. Two ML models, multi-layer perceptron and support-vector machine, were used to learn the relationship between the number of signal events and the parameters of CMSSM. It was found that the accuracy of predicting the likelihood function can reach several percent with 2000 training samples.

		\item {\bf Learning the exclusion of new physics models at the LHC using a decision tree approach.}
		Exploring new physics is one of the key research topics at the LHC. In order to test whether any particular set of parameter points of the new physics model have been excluded by the LHC experiment, a lot of time and computational resources need to be spent to generate simulated events on the LHC. Also the scattering cross section calculation, detector simulation, event reconstruction, and event selection on hundreds of signal intervals in ATLAS and CMS experiments need to performed. So it is a challenging problem. In the project BSM-AI~\cite{Caron:2016hib}, ML models are used to predict whether a new physics model has been excluded by the LHC experiments based on the input model parameters in milliseconds. The first implementation case is SUSY-AI. At present, SUSY-AI model can reproduce the exclusion region of ATLAS with an accuracy of 93\% in the 19-dimensional parameter space of pMSSM. For CMSSM and NMSSM, the prediction accuracy is also very good. The establishment of these models will help to solve the problem of recasting the LHC experiments when testing any new physics model.

		\item {\bf Using neural networks to learn likelihood functions to accelerate bayesian analysis of parameter space.}
		In~\cite{Graff:2011gv} an algorithm for fast Bayesian analysis is proposed, which is called blind accelerated multimodal Bayesian inference (BAMBI). It combines the advantages of nested sampling and artificial neural network, and improves the MultiNest method with neural network to learn the likelihood function. For the likelihood function which is very complicated to calculate, this algorithm can use the neural network to calculate the likelihood function quickly and thus greatly improve the efficient of Bayesian analysis. This algorithm has been used in the Bayesian analysis of the parameter space of cosmological model. The BAMBI method was found to have a great acceleration in various physics models and data.

		In another work~\cite{Bechtle:2017vyu} a tool named SCYNet is proposed to verify the supersymmetric model based on the LHC data. This tool uses neural networks as regressors to quickly estimate likelihood function ratios. The work designs two neural network models. The first one uses the 11-dimensional model parameters of pMSSM as input in the training. It can estimate the likelihood function ratio in milliseconds, which is then used for fast scanning of pMSSM model parameters. The second one uses designed model-independent features as inputs, such as the energy and particle multiplicity derived from a given new physics model. Although such an approach requires some time to calculate model-independent input parameters, the corresponding neural network obtained is more universal. So the neural network are trained only once but can be used to predict the likelihood function ratios of various new physics models at the LHC.

		\item {\bf Using neural network to determine the parameters of new physics model.}		
		Most of the researches studied the mapping of new physics model parameters to experimental observables. However, few researchers have inversely studied the mapping of the physical observables at the LHC to the parameters of the new physics model, aiming to directly determine the unknown physical parameters in the minimal supersymmetric model. In~\cite{Bornhauser:2013aya} a multi-layer perceptron is trained with 84 physical observables at the 14 TeV LHC as inputs and taking the parameters of supersymmetric model as the target outputs. It was found that if the collider luminosity is 10~fb$^{-1}$, the CMSSM model parameters $M_0$ and $M_{1/2}$ can be determined reliably, only with 1\% error. If the collider luminosity reaches 500~fb$^{-1}$, the model parameters $\tan\beta$ and $A_0$ can also be very accurately determined. As a contrast, the traditional method of minimizing $\chi^2$ gives relatively poor results.

		\item {\bf Active incremental learning for parameter space analysis.}
		We recently proposed a novel active incremental learning analysis framework to achieve more fast and reliable exploration of new physics parameter spaces~\cite{Ren:2017ymm}. As a proof-of-concept, we applied this approach to several benchmark models and two real physical models. It can be found that such a method can significantly reduce the computational cost and ensure the discovery of all survived regions.
	\end{itemize}

	\subsubsection{Jet tagging}

	The massive particles produced in a collider with high center-of-mass energy usually have high velocity. Their hadronic decay products are approximately collinear so that the jets formed by these products overlap with each other. Determining whether a jet object comes from a single light particle or from the decays of a heavy particle is an important task in the data analysis of the collider. Traditional techniques rely on the expert constructed distribution features of energy deposited in the calorimeter cells. But the complexity of the data makes machine learning methods more effective than human works ~\cite{Cogan:2014oua, deOliveira:2015xxd, Komiske:2016rsd, Baldi:2016fql, Almeida:2015jua, Kasieczka:2017nvn, Barnard:2016qma, Pearkes:2017hku, Butter:2017cot, Louppe:2017ipp, Cheng:2017rdo, Henrion2017, Aguilar-Saavedra:2017rzt, Datta:2017rhs, Chang:2017kvc, Datta:2017lxt, deOliveira:2017pjk, Paganini:2017hrr, Larkoski:2017jix, Bhattacherjee:2019fpt}.
	
	\begin{itemize}
		\item {\bf Tagging with jet images.}		
		In~\cite{Cogan:2014oua} the concept of jet image is proposed. The detector is regarded as a camera, and the energy distribution of jet in calorimeters is regarded as a two-dimensional digital image, where each pixel corresponds to the energy deposited in a specific cell of the calorimeter. Thus the jet tagging task becomes a standard pattern recognition problem and the jet data can be handled using machine learning algorithms for image classification. Taking differentiation of the hadronic $W$ boson decay from the jet generated by a quark or gluon as an example, this work classifies jet images using Fisher classification analysis in machine learning. With Monte Carlo simulation, it is found that this method can achieve better jet discrimination ability than traditional jet substructure methods, and gives a deeper insight into the internal structure of jet.

		In~\cite{deOliveira:2015xxd} the CNN is further used to investigate the hadron decay of the boosted $W$ boson. This deep learning technique can go beyond the traditional approach of building jet features based on physical intuition. At the same time, this work also visually presents learned jet features to explain why CNN can have better performance, which in turn can be used to further understand the physics in jet.

		In~\cite{Komiske:2016rsd} the CNN is used to distinguish quark and gluon jets, and obtained a better tagging performance than expert-constructed jet observables. This work not only takes the energy distribution of jet in the calorimeter as an image, but also uses different image channels to respectively represent the transverse momentum of charged particles, the transverse momentum of neutral particles and the number of charged particles in the calorimeter cell. The results show that the CNN can achieve or exceed the traditional jet tagging method using jet observables. At the same time, similar to traditional jet observables, the CNN is not sensitive to the quark and gluon jets generated by different event generators. Therefore, CNN can be used to extract reliable physical information from imperfect collider simulation data.

		In~\cite{Baldi:2016fql} the learning is performed on the data using a DNN with both local connection layers and full connection layers. It is found that the DNN can reach or even exceed the performance of the best traditional algorithm without the jet observables built by experts. Moreover, the DNN maintains its advantage even after considering the effects of pileup collisions.

		In~\cite{Almeida:2015jua} the NN was trained using a large number of Monte Carlo samples of boosted top jets and light quark or gluon jets. The results show that the NN marking method has very good classification performance. When jet's $p_T$ is in the range of $1100-1200$~GeV, such an approach can achieve a top tagging efficiency of 60\% at a mis-tagging error of 4\%. This work also discussed the most important jet features that NN learned, and the relationships between these jet features and the known hand-built jet observables. Furthermore, the work~\cite{Kasieczka:2017nvn} compared the CNN with the QCD based multi-variable top-tagging. It was found that they have similar tagging performance.

		Although DNN based machine learning techniques have good application prospect in jet classification, these ML models require a large number of training samples from Monte Carlo event generator. The data distribution of these generated training samples is not completely consistent with experimental data at the colliders, which makes ML models biased. Therefore, the work~\cite{Barnard:2016qma} investigated the performance of DNN classifiers with different parton shower phenomenological models and the detector-related effects. It was found that the background suppression of ML models trained with samples from different event generators can reach 50\% for the same signal selection efficiency. In addition, this work also studied the applicability of DNN to different jet transverse momentum scales. These studies will help enhance our understanding of jet features learned from DNN.

		\item {\bf Jet classification using particle information.}
		Although jet is regarded as an image in the calorimeter and the performance of detecting jet substructure can be improved greatly by combining the advantages of DNN in image classification, such methods also encounter some problems, such as sparsity of jet images. In addition, building jet images by pixelation or building advanced jet features can result in precision loss. In addition to the method of classifying jet images, in~\cite{Pearkes:2017hku} the input of the DNN are constructed using the four-momenta of the particles in jet. In this way, for reconstructed top jet with transverse momentum between 600~GeV and 2500~GeV, the suppression rate of QCD background jet can reach 45\% when the tagging efficiency of top jet is 50\%. The calculations also showed that this method is not sensitive to the interference from the LHC Run 2 pileup collisions.

		The study in~\cite{Butter:2017cot} used the Lorentz vector and the Minkowski metric to construct the features of the final-state particles as the input of DNN, and then constructed a special DNN layer. This approach allows us to identify top jet not only from the data in the calorimeter, but also from the particle track information. It was found that this DNN can further improve the tagging performance, compared with other ML methods such as image classification.

		The study in~\cite{Louppe:2017ipp} performed jet tagging using recursive neural networks (RNN). This method utilizes the similarity between jet clustering process and the structure of natural language, by considering the four-momenta of the final-state particles in jet as the words in natural language, and regarding the history of jet clustering using sequential combination clustering algorithm as the analysis process of the grammatical structure of natural language. The jet clustering process forms a tree structure and each jet has its own unique tree structure. The RNN is much suitable for processing the data of this kind of tree structure, so that the four-momenta of the final-state particles in jet can be directly used, and it works with any number of final-state particles. The results showed that this method can construct a vector representation of the input jet, and it has a higher data utilization efficiency and a better prediction accuracy than the ML method based on jet image. This work further extends the method to the classification of whole events, thus implemented the event classification directly from the data of the stable final-state particles of event at the LHC.

		In~\cite{Cheng:2017rdo} the RNN is further applied to the tagging of quark and gluon jet. It was found that the gluon jet suppression rate of the RNN is only a few percent higher than that of the BDT. Therefore, some relevant factors affecting the tagging performance of RNN are studied in this work. It was found that even if the information of momentum and position of the final-state particles are not given, the tagging efficiency can be well enough based on the tree structure formed by jet clustering history. Therefore, the main classification features of quark and gluon jet have already been reflected in the tree structure formed by jet clustering history. In addition, preliminary results on the tagging of up and down quarks are also given in this work.

		In~\cite{Henrion2017} the final-state particles in jet are represented as nodes in the graph, and jet is tagged using a convolutional MPNN. This work studied three kinds of MPNN. The first one has an learnable adjacency matrix, the second one has an learnable symmetric adjacency matrix, and the third one has an unit adjacency matrix with aggregated hidden state. It was found that the MPNN with learnable adjacency matrix and two message passing iterations gets the best jet tagging performance, compared with other NN models.

		\item {\bf Use deep learning to build model-independent jet tagging algorithm.}	
		New physics particles produced at the LHC typically have large boost, especially if they come from heavy particles. The signal generated by such a boosted particle will be a fat massive jet if it decays hadronically. Such a jet usually has a large QCD background. Although both the jet substructure method and the ML method can identify specific boosted jets from the QCD background, there is no jet tagging method that can be applied to all types of jets. In~\cite{Aguilar-Saavedra:2017rzt} a model-independent jet tagging method using $N$-subjettiness as NN input is described. It could identify various hadronical decays of boosted jets from the QCD background. The results showed that this method can be used to identify fat jet generated by any new physics model at the collider by $2-8$ times than the performance of traditional methods.

		\item {\bf  Using deep learning techniques to construct jet physical observables.}
		ML techniques are increasingly used in data analysis at the LHC. In particular, a large amount of ML work used image recognition, natural language processing and other algorithms for the identification of jets generated by different particles. However, these studies did not interpret the machines learned features that enable them to give a better tagging performance than the traditional manually constructed jet observables. Therefore, due to the black box nature, the ML is still widely criticized in the HEP phenomenological research.

		Take the identification of the hadron decay of boosted $Z$ boson from the QCD jet background as an example to understand what ML models actually learned. In~\cite{Datta:2017rhs} the learned features were mapped to some known manually constructed jet observables. The results showed that the main features that distinguish $Z$ jet from QCD jet are the jet observables in the phase space of four-body final-states.

		In~\cite{Chang:2017kvc} the data planing method was used to identify which combinations of expert constructed jet observables are able to distinguish the signal from the background. The weights of input features were introduced to suppress some input features and the NN was trained on the new data. If the classification ability of the NN decreases significantly after the removal of some input features, it indicates that the removed feature has an important influence on the jet tagging performance and thus it is a main features that distinguishes the signal from the background. In addition, using a toy model and a large mass resonance model at the LHC, this work also showed that the planning method can also be used to study whether the classification boundary of signals and backgrounds is linear or non-linear.

		In addition to studying what classification features ML models learned, the study in~\cite{Datta:2017lxt} used ML as an auxiliary to construct new jet tagging observables. This method first grouped jet observables according to particle phase space, and then measured how many jet observalbes should be used to saturate the classification ability of the ML model. If the $N$-body jet observables saturate the performance, the new jet observables can be constructed according to these $N$-body jet observables. In this work, a general form with adjustable parameters is adopted as the newly constructed jet observable. By adjusting the parameters in the new observable, it will get a classification performance approaching the ML model. Taking the distinction between $H \to b\bar{b}$ and $g \to b\bar{b}$ as an example, a new jet observable is constructed, which shows great classification performance than the traditional jet observable constructed from theory.

		\item {\bf Jet image generation.}
		Studying new physics at the LHC requires Monte Carlo simulation of new physics signals and backgrounds, so as to give the expectation of experimental data. Especially in experiments, the number of simulated events used for data analysis is extremely large, and it takes a lot of time and computing resources to generate the required simulated events using existing algorithms. When it comes to exactly calculating how energetic particles interact with materials in the detector, such calculations can be extremely time-consuming.

		Applying the generation model in the field of ML to the simulation of the physical process at the collider, the study in \cite{deOliveira:2017pjk} proposed to use the generative adversarial network (GAN) to artificially generate jet images in the calorimeter. A large number of simulated high energy collider events are used to train a kind of GAN so as to to generate realistic jet energy distribution. It was found that the pixel brightness of jet images generated by the GAN can span many orders and give the expected low-dimensional physical observables like reconstructed jet mass and $n$-subjettiness. At the same time, the limitations of this method and an empirical validation of the image quality is also presented. If this method can be further improved, it can be used to simulate HEP events quickly.

		In~\cite{Paganini:2017hrr} the feasibility of deep generation model in high-quality, fast and electromagnetic calorimeter simulation is further studied. It was found that although it is difficult to accurately simulate the whole phase space with deep generation model, this method can reproduce many simulation properties of the detector and can accelerate the simulation of calorimeter by 100,000 times.

	\end{itemize}

	\subsubsection{Collider event classification}

	Particle collisions in the high energy collider are the most effective experimental approach to generate new particles. It is always a critical task to identify signals from background events in the search for rare new particles. This is a major application of ML techniques in the experimental HEP research. Standard data analysis methods mainly use traditional or shallow learning methods, but these ML models have limited ability to learn complex nonlinear functions in the input data and rely on manually constructed nonlinear event features. A large number of traditional techniques show similar performance. The development of deep learning techniques make it easy to learn complex and non-linear functions to better distinguish signal events from background events at the collider.

	\begin{itemize}
		\item {\bf Event classification with DNN.}
		In~\cite{Baldi:2014kfa} two groups of new physics signals were used for validation. It was found that the deep learning method not only needs no expert constructed event features, but also improves the performance of classifying signal and background events by 8\% compared with the best existing methods.

		The Higgs boson provides the masses for the fundamental fermions. Although the measurements at the LHC are still consistent with the standard model, the traditional analytical techniques used in the experiments cannot enhance the significance to $5\sigma$ without adding additional data. Deep learning technique provides the ability to automatically learn complex data. So, in~\cite{Baldi:2014pta} the deep learning technique is applied to the detection of Higgs decaying to $\tau^+ \tau^-$ at the LHC. The hyperparameters of DNN are optimized by Bayesian optimization algorithm. The detection significance is 25\% higher than the existing traditional analysis techniques.

		The article~\cite{Demir:2018iqo} used DNN to study the signal of the light charged Higgs particle produced by the top quark decay at the LHC. The difficulty of finding this signal comes mainly from the large $W$ boson background. Taking the low level kinematic features of events with the expert constructed event features as input, respectively, it was found that the DNN not only needs no expert constructed event features as inputs, but also has better classification performance than the shallow NN. It also showed that DNN can automatically learn the information of manually constructed event features from data. In addition, increasing the number of neurons in the DNN will lead to the increase of the complexity of the NN model, so it cannot effectively improve the classification performance.

		In the framework of the standard model efficient field theory, the article~\cite{DHondt:2018cww} studies the observability of four-fermion operators containing heavy quarks at the LHC through the $t\bar{t}b\bar{b}$ process. In this work, the kinematic features of the events and the NN model were used to optimize the detection sensitivity of the signals at the LHC. The NN was used to classify the top quark events with various helicity. It provides a new direction for testing effective operators and can improve the measurement of top quark interactions at the LHC.

		\item {\bf Event classification with CNN.}		
		The work~\cite{Madrazo:2017qgh} studied the application of CNN to event classification in high energy particle collision. It proposed to represent the jet types and momentum information of the final-state particles in the image in the $\eta$ versus $\phi$ plane, which are used to train CNN for event classification. By testing a set of simulated data of CMS detector for the singal and background of leptonic decays of $t \bar{t}$ process at 7~TeV LHC, it was found that the CNN and DNN have similar classification performance.

		The Higgs particle production with large $p_T$ at the LHC can be used to study the internal interaction structure of the $gg \to H$ loops. The decay of Higgs particle to $b\bar{b}$ has the largest branching ratio, which is conducive to the statistical analysis of the event. Although this process has a large QCD background, using jet substructure technique can improve the observability of $gg \to H \to b\bar{b}$ at the LHC. To further improve the detection sensitivity of this process, a two-path CNN was constructed in~\cite{Lin:2018cin}. One path in CNN takes the jet information as input, and the other takes the features of the whole event as input. This NN greatly improved the detection sensitivity of the Higgs particle at the LHC, which not only helps to detect the Higgs particle with large $p_T$ in the standard model, but also can detect the Higgs particle production in new physics models. Unlike the hadronic decays of other massive particles, the $b$-tagging in the two sub-jets of fat jet generated by the deacy of Higgs boson with large $p_T$ can suppress almost all jet backgrounds.

		\item {\bf Event classification with MPNN.}
		In~\cite{Abdughani:2018wrw,Ren:2019xhp} we propose to represent events as graphs and use MPNN to search for the stop pair production and $Ht\bar{t}$ production at the LHC. We found that the signal and background events can be efficiently discriminated by the patterns of event graphs. Such an approach can thus improve the current LHC detection sensitivity for the stops and helps to reveal the CP nature of the top-Higgs interaction. In the following section, we will give a expanded description for these applications.

		\item {\bf Comparison of various NN.}
		In~\cite{Nguyen:2018ugw} the performance of DNN, CNN and RecNN in real-time event classification at the LHC are compared. This work considered various data representations of collider events and trained corresponding deep learning models as collider event classifiers. In the comparison, the original event data and the advanced event features constructed by human were used respectively. By analyzing some specific data, it was found that the background selection efficiency can be effectively reduced by one order of magnitude under the condition that the signal selection efficiency is 99\%. If these ML models are applied to the online event selection in LHC collider experiments, the storage of a large number of irrelevant background events for subsequent analysis can be reduced, thus significantly reducing the operation cost of detectors and improving the efficiency of data storage.

	\end{itemize}

	\subsubsection{Other miscellaneous applications}

	\begin{itemize}
		\item {\bf Pileup subtraction.}		
		A collision in the LHC collider typically involves many protons, which forms pileup. In addition to the main collisions, other soft collisions produce noise in the detector. The pileup collisions contaminate the main collision event and make it difficult to reconstruct the jets. The work in~\cite{Kong:2015phl} used ML technique to find important data features to accurately reconstruct the energy of final jets. It examined the performance of various ML models, including linear regression, SVM and decision trees. It is found that the linear regression model gives the best results with predictive data features as input. Compared with the benchmark model, the ML method greatly improves the performance of jet reconstruction.

		The study in~\cite{Komiske:2017ubm} used CNN to eliminate the effects of pileup. For each collider event, the energy distribution of charged particles generated from the primary colliding vertex and from the pileup as well as all neutral particles generated from collisions in the detector are taken as the input of CNN, and the energy distribution of all particles generated from the primary collision vertex in the detector is taken as the output of CNN. Such a method is called PUMML in the work. It is found that this algorithm can greatly suppress the energy noise generated by th pileup in the detector, so as to improve the reconstruction accuracy of the final jet observables. In addition, the stability of the algorithm is tested.

		\item {\bf Parton distribution function fitting.}		
		The greatest challenge of the parton distribution function (PDF) is to estimate the uncertainty of the combined PDF from a single PDF. Since 2014 several approaches have been proposed to tackle this problem. The literature~\cite{Carrazza:2016sgh} summarized the recommended strategies in PDF4LHC15 and a new clustering method using ML is discussed.
		
		\item {\bf Phase space integral of final-state particles.}		
		The literature~\cite{Bendavid:2017zhk} proposed the Monte Carlo integration method using the generative decision tree model and the generative DNN, respectively. Taking integrand that cannot be factored as test cases, it showed that both ML methods can greatly improve the accuracy of integration, compared with the existing integration methods, using the same number of integrand evaluations. If the stability and performance of this method are further validated, combing this integration method with the scattering amplitude computation can improve the efficiency of Monte Carlo based collider event generation.

	\end{itemize}

	\subsection{Active incremental learning in parameter scan}

	The research on parameter space of new physics models is an important part in the process of discovering new physics. However, with the constraints of experimental data, it is very time-consuming and difficult to fully explore the parameter space of new physical models and find all the parameter regions that meet the experimental constraints. Especially when the parameter space of a new physical model is high dimensional, or there are several independent survived parameter regions, the work of parameter space scanning becomes more challenging. Therefore, we proposed a new method of using active incremental learning to accelerate the parameter space scanning of new physical models, which we call machine learning scan (MLS)~\cite{Ren:2017ymm}. In this method, the parameter space is explored by iterative and incremental learning the historical data to actively guide the subsequent parameter sampling. Firstly, active learning can greatly improve the efficiency of parameter sampling. Secondly, through incremental learning, the ML model can be continuously optimized. In addition, the accuracy of the learned physical observables can be greatly improved by using the DNN. All the sampled parameter points, including some additional random parameter points, are used for the accurate reconstruction of the global parameter space. As validation, we tested this MLS method in several toy models and compared its performance with other methods. Further, the MLS method is applied to the parameter space exploration of two given new physics models. We found that the MLS method can greatly reduce the computing cost, and at the same time has a better ability to find the multiple independent survived parameter regions.
	
	Note that some ML related works have been done using ML to accelerate parameter space analysis for new physics models~\cite{Bridges:2010de, Buckley:2011kc, Caron:2016hib, Bechtle:2017vyu, Bornhauser:2013aya}. However, most of them use off-line learning to train ML models to approximate physical observables or likelihood functions. Because the performances of these methods depends heavily on the complexity of the ML models and the quality of initial training samples, it is difficult for them to obtain high local precision. BAMBI~\cite{Graff:2011gv} is a method of online learning, which uses newly collected samples to incrementally train a shallow NN. The role of NN in this method, however, is only to improve the local precision of the likelihood function calculation in the MultiNest algorithm~\cite{Feroz:2008xx}; it does not improve the MultiNest algorithm in finding the independent survived regions in the whole parameter space. In the following we provide a detailed description of the MLS method based on our recent work~\cite{Ren:2017ymm}.

	\subsubsection{Description of the method}

	{\bf Approximation of the likelihood function:}
	Likelihood function is widely used to measure the quality of model parameter $\bm{x}$ in interpreting experimental data. We usually use the definition
	\begin{equation}
		L(\bm{x}) = \prod_i L_i(O_i(\bm{x}); O_i^*, \sigma_i^*) ,
	\end{equation}
	where $O_i(\bm{x})$ is the theoretical prediction of the physical observable $O_i$ at model parameter point $\bm{x}$ and $O_i^*$ is the corresponding experimental value with an uncertainty $\sigma_i^*$. To simplify our analysis, we ignored the correlation between experimental data.

	For a new physics model with many free parameters, in most part of the parameter space the likelihood $L(\bm{x})$ is close to zero. Moreover, it is possible to have a large likelihood value only on or near a very thin "surface". In addition, a new physics model may have multiple survived parameter regions, which are usually independent and far apart in the parameter space. Therefore, using a limited number of random parameter samples, it is insufficient or difficult to reconstruct a likelihood function accurately through a ML model.

	Note that compared with the likelihood function, the physical observable functions usually change very slowly with the change of parameter values, which makes them more suitable for ML. Using a sufficient number of parameter samples, the trained ML models can reconstruct the physical observable functions with high accuracy. Therefore, we do not directly train the ML model to fit the likelihood function, but train the ML model to learn the physical observable functions.

	We use the symbol $M_i$ to represent the ML model of physical observable, then
		$O_i(\bm{x}) \approx \hat{O}_i(\bm{x}) = M_i(\bm{x})$,
	with $O_i(\bm{x})$ representing the theoretical predictions of the new physics model on the physical observables when parameters take value of $\bm{x}$, $\hat{O}_i(\bm{x})$ being the output given by the ML model $M_i$. If we use the physical packages to calculate the physical observables $O_i(\bm{x})$, it can be rather time consuming. Usually it may take seconds to hours to calculate all physical observables for a single parameter point of a new physics model, or even several days if the collider simulation is needed. However, using the approximate calculations from the ML models, all physical observables for millions of parameter points may be calculated in about one second. Since the ML model has a powerful generalization capability, the likelihood function can be further approximated by
	$\prod_i L_i(\hat{O}_i(\bm{x}); O_i^*, \sigma_i^*)$.

	{\bf Active and incremental learning:}
	In order to explore the parameter space of the new physics model with ML, it is necessary to train the ML model so that it can reconstruct the likelihood function accurately. Usually, the ML model is trained only once, using the method described in the literature~\cite{Bridges:2010de, Buckley:2011kc, Caron:2016hib, Bechtle:2017vyu, Bornhauser:2013aya}. In order to achieve a good approximate accuracy, the only method is to collect a large amount of parameter sample data to train the ML model. However, generating a large amount of parameter sample data is a computing intensive work, because it is necessary to generate enough random parameter points and calculate the prediction of the physical observables for each parameter sample. Therefore, we adopt active and incremental learning strategy to improve or avoid such problems.

	\begin{figure}[t!]
		\centering
		\includegraphics[width=14cm]{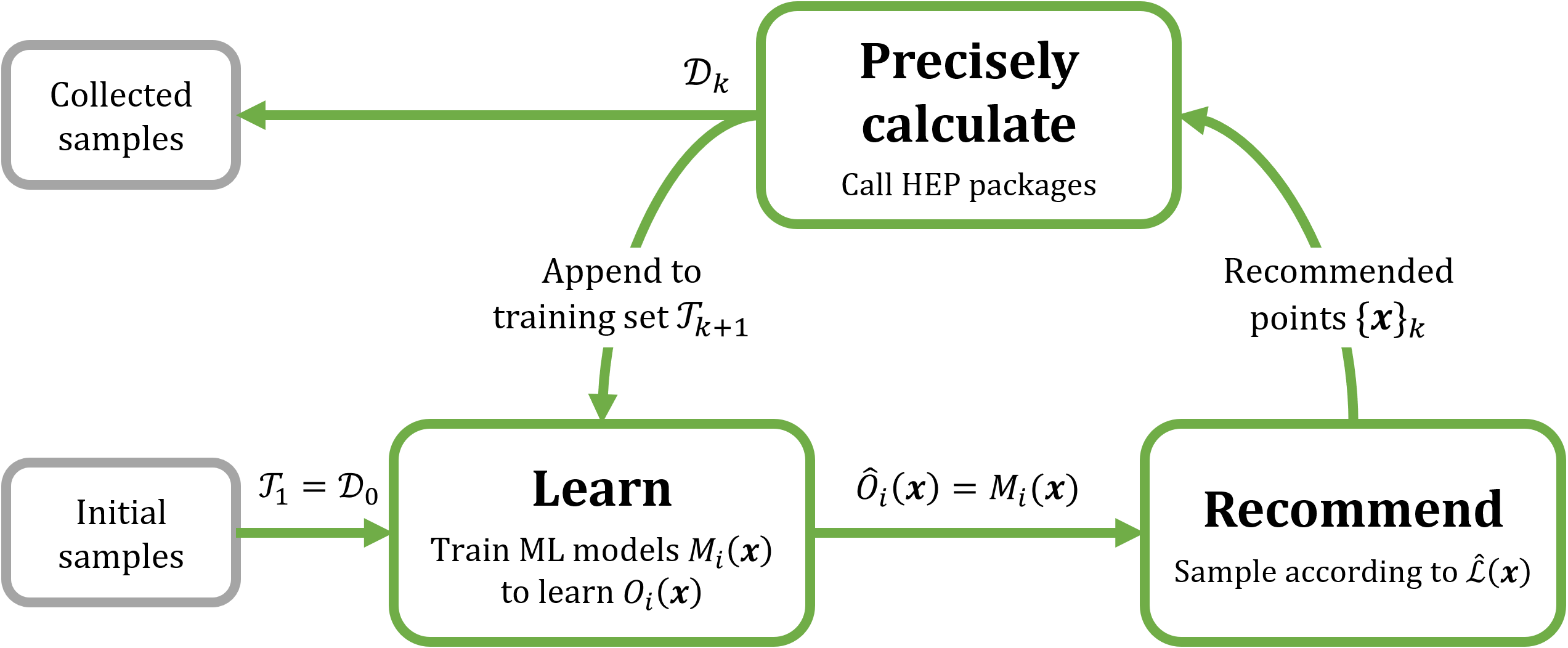}
		\caption{The flow diagram of active and incremental machine learning scan algorithm. This figure is taken from \cite{Ren:2017ymm}.}
		\label{Chap3/workflow}
	\end{figure}

	Fig.~\ref{Chap3/workflow} shows the overall workflow of our method. In the $k$-th iteration,
	we first use the collected parameter sample data $T_k$ to train the ML models to obtain the approximate function $\hat{O}_i$ of the physical observables, and further obtain the approximate function $\hat{L}$ of the likelihood. Then, according to the approximate likelihood function $\hat{L}$, this method actively identifies the important parameter regions and recommends some parameter points within these regions which are worthy of further exploitation. We then use the physical packages to calculate the exact values of physical observables for each of the recommended parameter sample. These data will be used as the newly collected sample data $D_k$ and appended to the training sample data set $T_k$: $T_{k+1} = T_k \cup D_k$. In the next iteration, $T_{k+1}$ will be used to incrementally train the ML models, in order to gradually improve the accuracy of the ML models in the reconstruction of physical observables and likelihood functions. The above process is repeated until a sufficient number of parameter samples are collected for subsequent physical analysis. In this way, the ML models will become more and more accurate based on more and more high-quality sample data. At the same time, with the continuous improvement of ML model accuracy, the parameter regions that need to be focused on will be constantly optimized, and the boundaries of these regions will become clearer and clearer. Finally, more and more parameter samples are collected in the survived parameter regions, and the ML model becomes very accurate in approximating the physical observables and likelihood functions within the survived parameter regions. In addition, the ML model also has a rather good reconstruction accuracy in the nonviable parameter regions. In this way, the sampling efficiency of parameter points will be greatly improved since there is only a small chance to sample the parameter points outside the survived parameter regions.

	The implementation details of our MLS method are as follows.
	\begin{enumerate}
		\item The initial training set $T_0$ usually has a small number of data samples. These parameter points are usually generated by random sampling. Of course, it can be much useful to add the previously existing data samples, which can further improve the efficiency of the parameter space analysis of the new physics model.
		
		\item MLS is a general framework and is not limited in the types and specific choices of ML models. By default, we choose the DNN because it has been proved to be powerful enough for data fitting~\cite{Cybenko1989, Hornik1991}. Of course, we can choose any ML model with sufficient representation capability and generalization ability.
		
		\item Because ML model can evaluate millions of parameter points in a very short time, we took the approximate likelihood function $\hat{L}$ as the target distribution and adopted simple rejection sampling to sample the "important" parameter regions. Of course, adding prior knowledge or experience about the parameter space of the new physics model into the sampling algorithm can further accelerate the sampling of the important parameter regions and further improve the quality of the sampled points.
		
		\item In order to find more important parameter regions, in addition to the points sampled according to the approximate likelihood function $\hat{L}$, it is also crucial to generate some random parameter points. If we need to encourage the algorithm to find more important parameter regions, we need to generate more random parameter samples. On the contrary, with fewer random parameter samples, the algorithm will focus on exploring the important parameter regions it already found.
		
		\item It is worth noting that because it is very time-consuming to calculate physical observables using physical packages, all the collected samples are used to train the ML model without the validation, testing, and regularization that normally required in machine learning. Using a subset of all the samples as training samples may result in the failure of ML models to learn some key features in the parameter space. Regularization the ML model will inevitably smooth the steep features in the parameter space to some extent, thus affecting the approximate accuracy of ML models near these regions. In ML, lack of validation, testing and regularization often leads to severe overfitting. Although overfitting is inevitable, this problem can be controlled in our method. Combined with the characteristics of active learning and incremental learning, if the ML model mistakenly learns the false important parameter regions, the algorithm will subsequently collect a certain number of data samples in these regions, which will be used to correct the ML models accordingly.
	\end{enumerate}

	For each of the recommended parameter points, we need to use the physical packages to accurately calculate the corresponding physical observables. Therefore, the data samples obtained by our method are all accurate. In addition, the final ML models are accurate enough to be used for the rapid calculation of physical observables in the subsequent parameter space analysis.

	We implemented our MLS framework in Python environment, in which the open source of deep learning framework PyTorch \cite{PyTorch} was used for the construction and training of DNN models. All calculations below are performed on machines with Intel Core i7-4930k and NVIDIA Titan XP. In order to test the performance and reliability of MLS method, we first applied it to some toy models. The first toy model is a simple one-dimensional model used to validate the MLS method. The second toy model has a number of independent survived parameter regions spaced far apart and is to used to test the capability of MLS to discover independently survived parameter regions. A third toy model is used to test the performance of the MLS method in high-dimensional parameter space. Finally, we applied the MLS method to the analysis of  parameter space of two new physics models.

	\subsubsection{One-dimensional toy model}
\begin{figure}[t!]
		\centering
		\includegraphics[width=13cm]{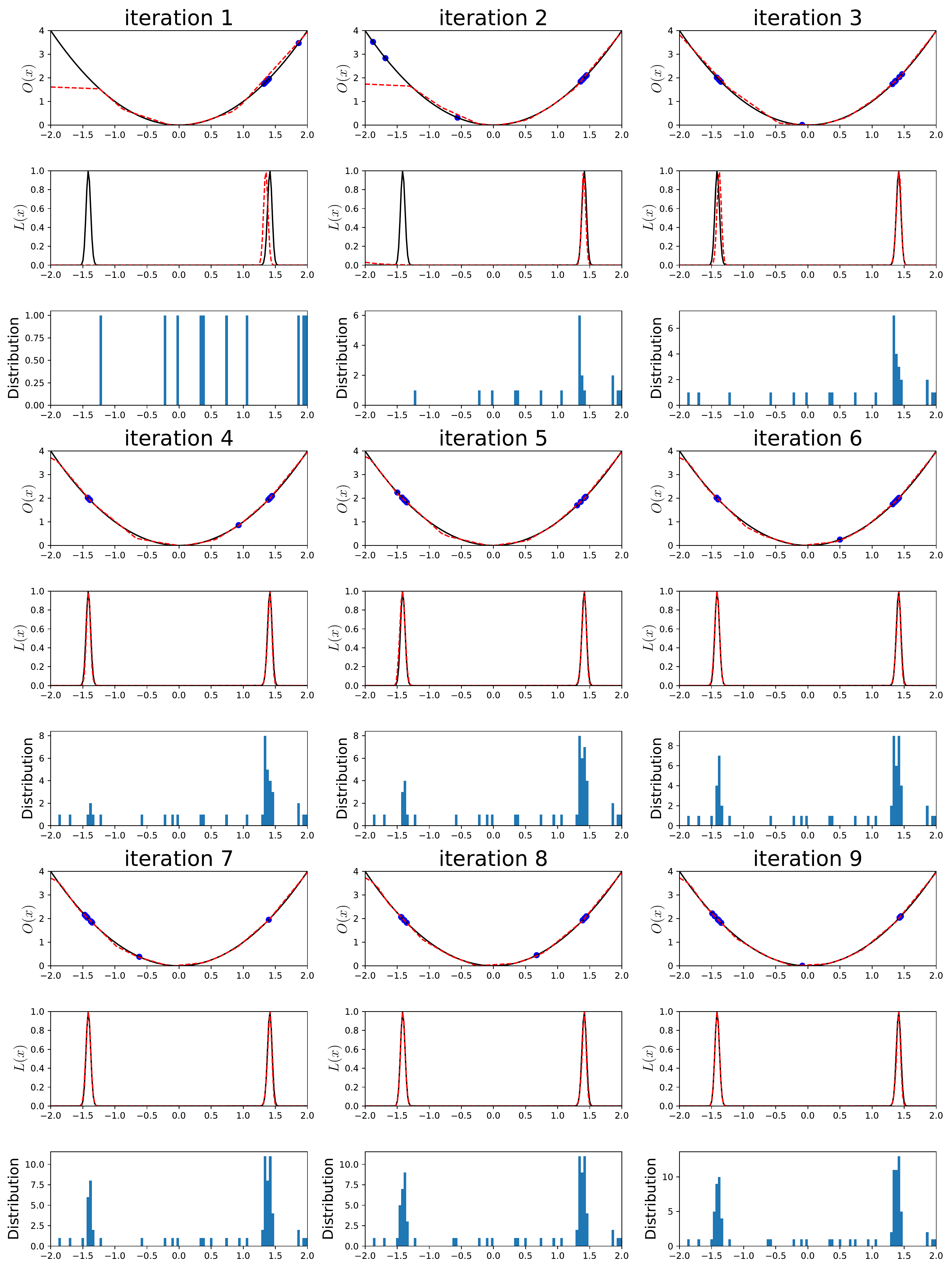}
		\caption{The first 9 iterations of MLS
	for the one-dimensional toy model. The black solid curves represent
	the theoretical values of the physical observable function and the likelihood function.
	The red dotted curves represent the physical observable function
	obtained from the deep neural network learning
	and the likelihood function obtained from the indirect reconstruction.
	The blue dots represent the newly generated parameter samples for each iteration.
	Histogram statistics of the distribution of all parameter samples in the
	parameter space are also given.}
		\label{Chap3/parabolic-1d}
	\end{figure}

	We first validate the MLS method on a simple one-dimensional toy model. In this model, there is only one physical observable
	\begin{equation}
		O(x) = x^2 \quad (-2 < x < 2) .
	\end{equation}
	Its likelihood function has a Gaussian form
	$ \exp [ - (O(x) - O^*)^2/2 (\sigma^*)^2]$.
	We assume that the experimental value $O^* = 2 $ with an uncertainty $\sigma^* = 0.1$. Fig.~\ref{Chap3/parabolic-1d} shows the above physical observable and likelihood functions. This toy model has the basic characteristics of the real new physics parameter space, that is, it only has a relatively large likelihood in a very narrow parameter region. Without some prior knowledge of the locations and shapes of these narrow survived parameter regions, it is difficult to analyze such parameter space directly using the traditional parameter space scanning methods.

	We use a deep feed-forward fully connected neural network to fit this physical observable function. This NN has one input layer, three hidden layers and one output layer. The input layer has only one neuron. Each hidden layer has 10 neurons, and each neuron is activated by the ReLU activation function. The output layer has only one linearly activated neuron. The standard ADAM optimization algorithm~\cite{KingmaB14} is used to train the learnable parameters in the neural network. In each iteration of the MLS algorithm, a fixed learning rate of 0.001 is used to train the NN up to 2000 epochs.

	First, we used 10 initial random samples to train the neural network. The first figure in Fig.~\ref{Chap3/parabolic-1d} shows the physical observable function learned by the NN and the reconstructed likelihood function. For each subsequent iterations, the MLS algorithm samples 9 parameter points according to the approximate likelihood function $\hat{L}(x)$, and generates one random parameter point at the same time. Fig.~\ref{Chap3/parabolic-1d} gives the physical observable function and the reconstructed likelihood function learned from the ML model after 9 iterations. After 9 iterations, the MLS algorithm generated a total of 90 parameter points (plus an initial sample of 10 random parameter points, a total of 100 parameter points). In the figure, we give the parameter point samples used to train the ML model in each iteration and the histogram statistics of the newly generated parameter point samples in the parameter space. Through the operation process of MLS algorithm, we can see clearly that with the constant improvement of the prediction accuracy of the ML model, the MLS algorithm used the clues provided by the approximate physical observable function to quickly find the two narrow parameter regions with large likelihood function values and then heavily samples in and around these regions. At the same time, as more and more parameter points are concentrated in the survived parameter regions, the prediction accuracy of the ML model in the survived parameter regions is rapidly improved.

	\begin{figure}[t!]
		\centering
		\includegraphics[width=12cm]{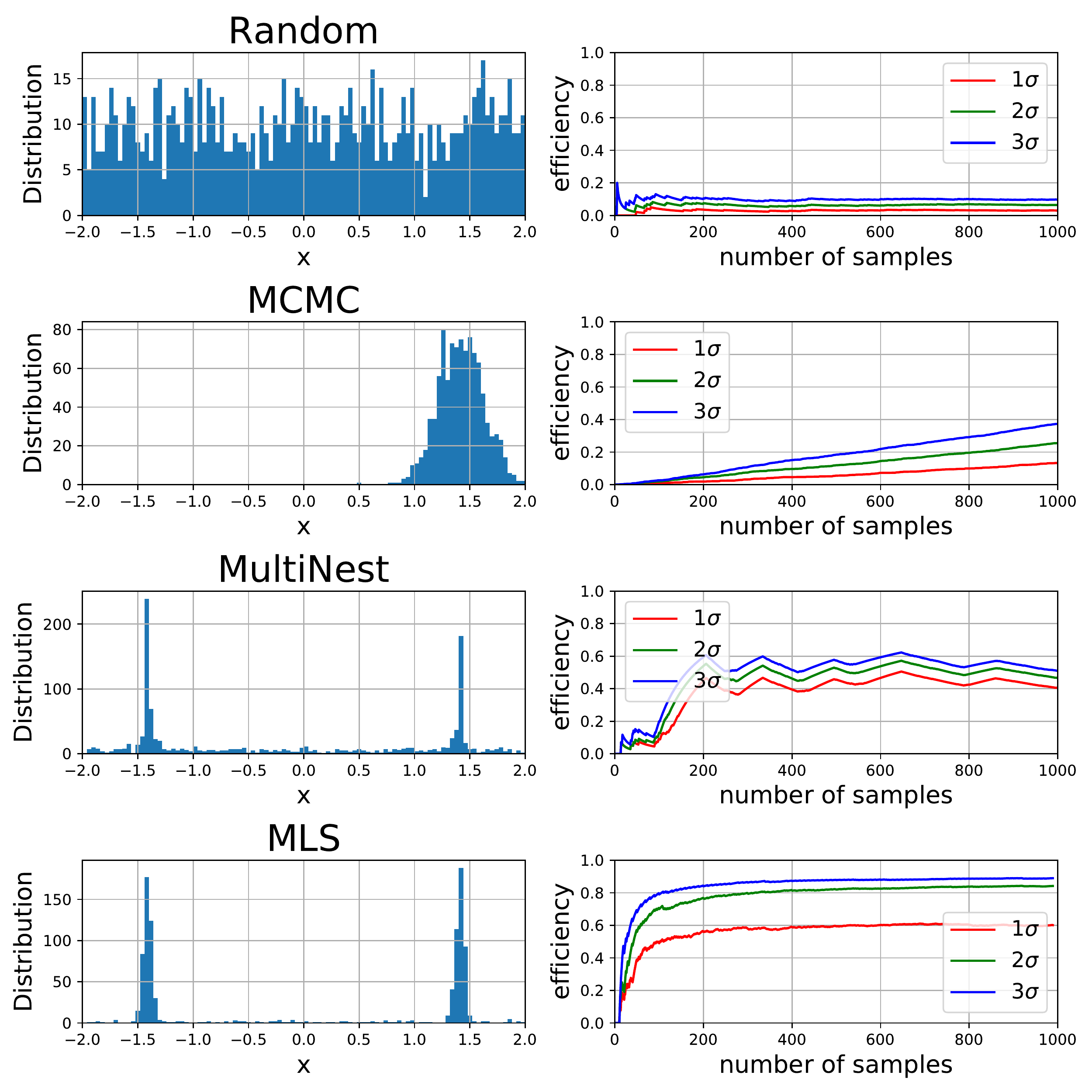}
		\caption{The parameter space analysis
	                 and the sampling efficiency of parameter points of each scan method
	                 for the one-dimensional toy model.}
		\label{Chap3/parabolic-1d-compare}
	\end{figure}

	In addition, we calculated the sampling efficiency of parameter points as shown in Fig.~\ref{Chap3/parabolic-1d-compare}. We measure the sampling efficiency by the proportion of the parameter points in the survived parameter regions with different confidence levels ($1\sigma$, $2\sigma$, $3\sigma$). For comparison, we also used the random scan, the MCMC scan and the MultiNest scan to analyze the model parameter space. We can clearly see that the random scan algorithm is an aimless sampling algorithm and the sampling is uniform in the whole parameter space. In this way, only a few sample points from the random scan method can fall into the survived parameter regions. Therefore, the sampling efficiency of the random scan method is very low. For the MCMC algorithm, the Gaussian distribution is adopted as the proposal distribution for the MCMC sampling. The width of the Gaussian distribution is set to 1/20 of the range of the parameter space. It can be seen from the MCMC sampling results that although the MCMC algorithm has a relatively high sampling efficiency, the MCMC algorithm is trapped in one survived parameter region and cannot find another independent survived parameter region, which is a major shortcoming of the MCMC scan method. As the most commonly used method of parameter space analysis, MultiNest can find two independent parameter regions with high sampling efficiency, but its sampling efficiency is slowly going up and not stable, which means it needs more parameter samples and large amount of calculation. In contrast, our MLS method can quickly find two independent survived parameter regions and rapidly improve the sampling efficiency.

	\subsubsection{Two-dimensional egg-box toy model}
\begin{figure}[t!]
		\centering
		\includegraphics[width=14cm]{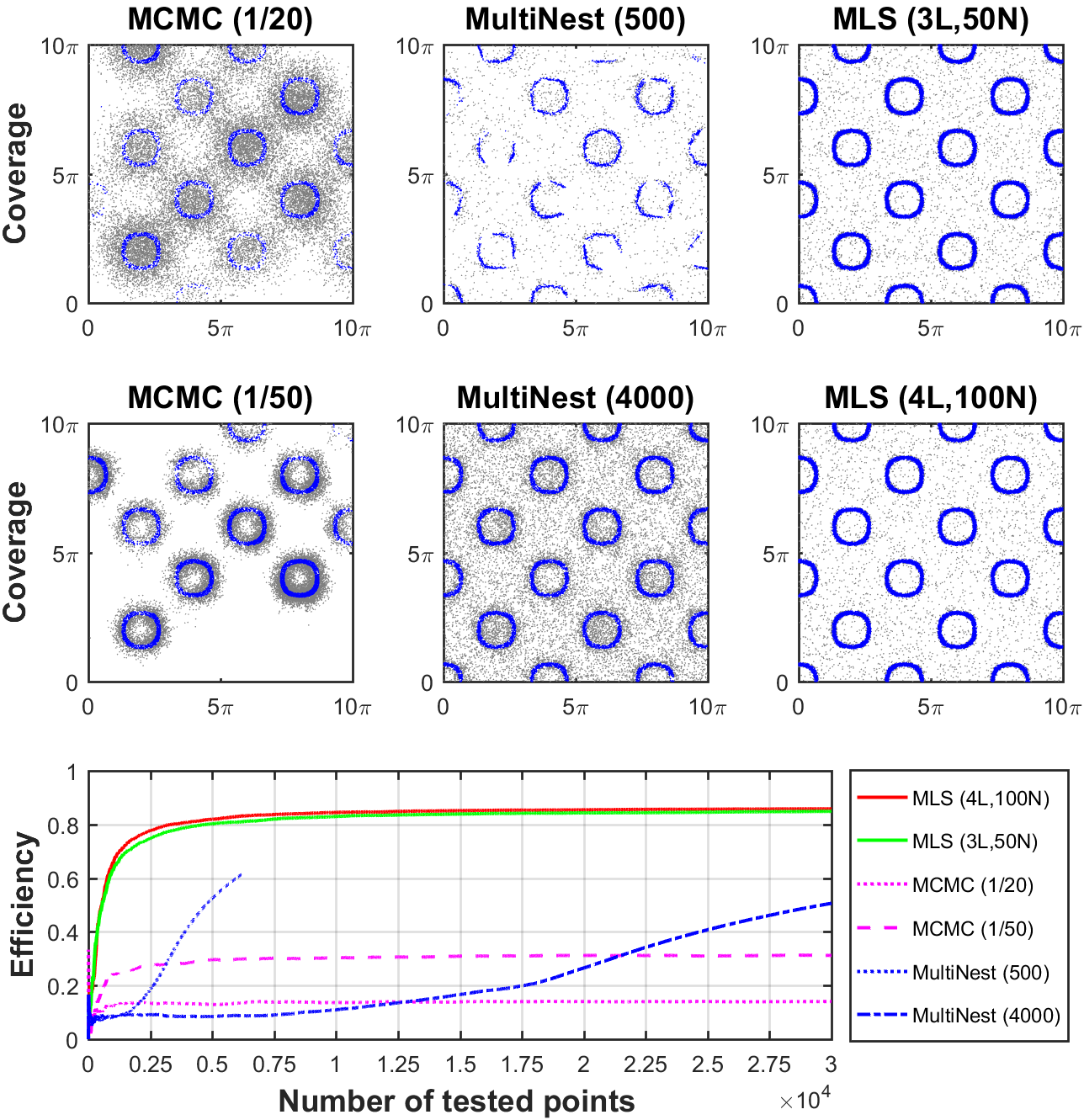}
		\caption{The parameter space analysis of the eggbox toy model:
	                 the sample coverage rates and the sampling efficiencies of
	                 machine learning scan, MCMC scan and MultiNest scan.
	The MultiNest scan used 500 live parameter points
	and other scans sampled $3\times 10^4$ parameter points.
	The corresponding algorithm parameters are marked in brackets.
	The blue points represent the parameter samples located in the survived parameter region
	($2 sigma$), and the gray points represent the non-survived parameter samples. This figure is taken from \cite{Ren:2017ymm}.}
		\label{Chap3/eggbox}
	\end{figure}

	Inspired by~\cite{Feroz:2008xx}, we tested the ability of MLS to find multiple independent survived parameter regions on a two-dimensional egg-box toy model. This model has two free parameters and one physical observable. The model parameters $x_1$and $x_2$ range from $[0, 10\pi]$. The physical observable takes the form
	\begin{equation}
		O(x_1, x_2) = \left( 2 + \cos\frac{x_1}{2} \cos\frac{x_2}{2} \right)^5 .
	\end{equation}
	As shown in Fig.~\ref{Chap3/eggbox}, because the shape of this physical observable is very similar to the shape of eggbox, this model is called the two-dimensional eggbox model. Further, we assume that the experimental value $O^* = 100 $ with an uncertainty $\sigma^* = 10$. We construct a Gaussian likelihood function
		$ \exp [-(O(x_1, x_2) - O^*)^2/2 (\sigma^*)^2]$.
	We need to find the parameter regions that satisfy the experimental data within the $2\sigma$ range. As shown in Fig.~\ref{Chap3/eggbox}, the target parameter regions are a series of independent and widely separated thin rings.

	Fig.~\ref{Chap3/eggbox} presents the results given by MLS, MCMC, MultiNest in analyzing the eggbox toy model. This figure shows the distribution of parameter samples and the sampling efficiency. For the MCMC method, we used 20 Markov chains and for each Markov chain the Gaussian proposal distribution is used to sample the parameter points. If the width of the proposal distribution is set to 1/20 of the range of the parameter space, the sampling of the parameter points will span several survived parameter regions, and the locations and shapes of the survived parameter regions cannot be found out accurately. Otherwise, if the width of the proposal distribution is set to 1/50 of the range of the parameter space, the sampling of the parameter points will be trapped in some local survived parameter regions, and other independent survived parameter regions with relatively long distances cannot be found out. Therefore, the sampling efficiency and sample distribution of MCMC method are very poor.

	For MultiNest method, using a small number (500) of live points, it is difficult to find all the survived regions of the parameter space. By fine tune this parameters, we found that the use of at least 4000 live points is needed to guarantee the finding of all survived parameter regions. However, in this case, its sampling efficiency will be greatly suppressed.

	Using our MLS method, we constructed a DNN to learn the physical observable function. This NN has one input layer, four hidden layers and one output layer. Each hidden layer has 100 neurons, and each neuron adopts the ReLU activation function. The output layer has a neuron with a linear output. We used the mean-square-error as the loss function and the standard ADAM optimization algorithm to train the learnable parameters in the NN model. The learning rate of the NN model was fixed at 0.001. Each iteration of the MLS algorithm trains the NN up to 1,000 epochs. In the experiments, we used 100 random parameter samples. Each iteration of the MLS algorithm generates 90 parameter points sampled according to the approximate likelihood function $\hat{L}(x_1, x_2)$, and 10 random parameter points. As a test, we also used another DNN with different hyperparameters. From Fig.~\ref{Chap3/eggbox} we find that the performance of MLS method has little dependence on the hyperparameters of DNN model. As long as the learning and generalization capability of the selected ML model are powerful enough, the sampling efficiency of the parameter points of MLS method is almost consistent, and it has the capability to consistently find all the independent survived parameter regions with relatively long distances.

	\begin{figure}[t!]
		\centering
		\includegraphics[width=14cm]{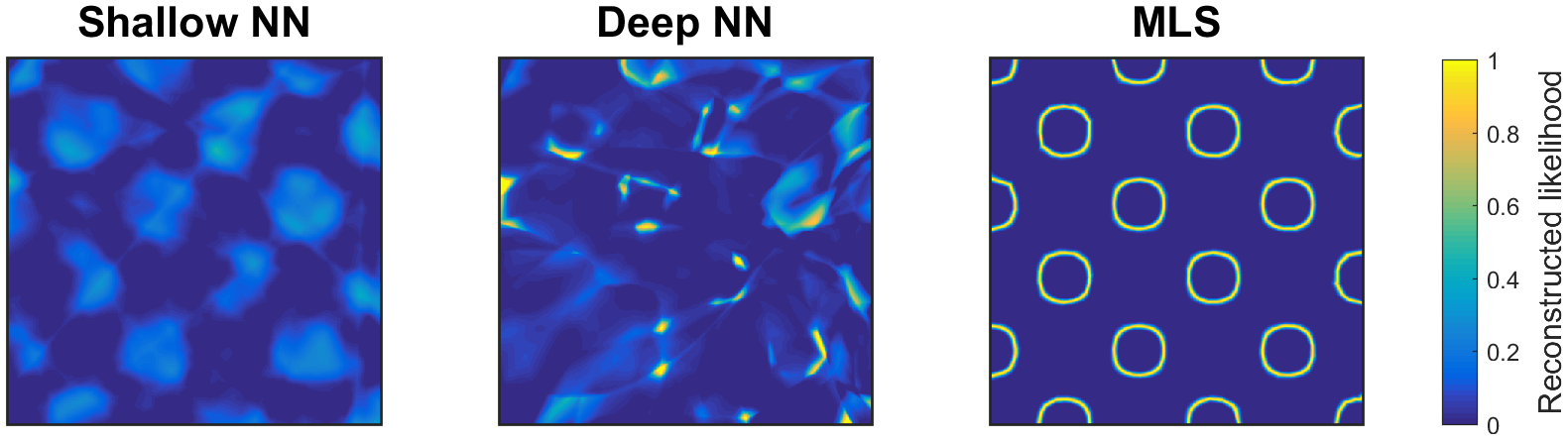}
		\caption{The likelihood functions reconstructed using the shallow neural network,
	the deep neural network and our machine learning scan algorithm.
	The first two methods use 2000 parameter samples to learn the target likelihood function directly.
	The machine learning scan algorithm runs until 2,000 parameter points are sampled.
	The colors in the figure represent the values of the reconstructed likelihood function. This figure is taken from \cite{Ren:2017ymm}.}
		\label{Chap3/eggbox-fit}
	\end{figure}

	In addition, we also carried out the experiments of learning the likelihood function directly with the shallow NN and the deep NN. The shallow NN has only a hidden layer of 2,000 neurons. The deep NN has 4 hidden layers, and each hidden layer has 100 neurons. We used a sample of 1800 random parameter points to train the two NN respectively. Unlike the MLS method, which trains the NN iteratively and incrementally, these two neural networks are trained only once. Additionally 200 random parameter points as the validation set were used for model selection, and the corresponding NN model parameters with the minimum loss on the validation set were selected. As shown in Fig.~\ref{Chap3/eggbox-fit}, we found that the shape of the likelihood function could not be accurately reconstructed by direct learning of the likelihood function, neither the shallow NN nor the deep NN. For comparison, the NN model in the MLS method can reconstruct the shape of the likelihood function very accurately when the MLS method samples 2000 parameter points. This is because the active parameter point sampling of MLS method can specifically improve the learning of the NN model on the physical observable in the parameter space, so as to better reconstruct the structure of the likelihood function.

	Therefore, by both theoretical analysis and practice, we can prove that if the likelihood function is directly fitted instead of indirectly reconstructed by learning the physical observable function, only if a sample parameter point is collected accidentally within a survived parameter region, can this survived parameter region be further discovered. So more parameter point samples are needed, which leads to the increase of computing cost.

	\subsubsection{High dimensional toy model}

	In order to test the performance of MLS in exploring high-dimensional parameter space, we built a high-dimensional quadratic toy model. This model has a $n$-dimensional parameter space and its unique physical observable is defined as
		$O(\bm{x}) = \bm{x}^T \bm{x} = \sum  x_i^2$.
	We adopt a gaussian likelihood function
		$L(\bm{x}) = \exp [ - (O(\bm{x}) - O^*)^2/2 (\sigma^*)^2]$
	where we assume the experimental value $O^* = 2 $ with an uncertainty $\sigma^* = 0.1$. Our task is to find the survived parameter regions that meet the experimental data within the range of $2 sigma$. According to the model, we can know that the survived parameter region is a hyperspherical shell in a high-dimensional space.

	We use a DNN to learn the physical observable function. This NN has one input layer, four hidden layers and one output layer. The input layer has $n$ neurons, corresponding to $n$ parameters of the toy model. Each hidden layer has 100 neurons, and each neuron adopts the ReLU activation function. The output layer has a neuron with a linear output. We use the mean-square-error as the loss function and the standard ADAM optimization algorithm to train the learnable parameters in the NN model. The learning rate of the NN model is fixed at 0.001. Each iteration of the MLS algorithm trains the NN up to 1,000 epochs.

	For the 2-dimensional model ($n=2 $), we used 20 initial random samples. Each iteration of MLS algorithm produces 18 samples according to the approximate likelihood function $\hat{L}(x_1, x_2)$, and 2 random samples. For the 7-dimensional model ($n=7 $), we use 100 initial samples. Each iteration of MLS algorithm produces 90 samples according to the approximate likelihood function $\hat{L}(x_1, x_2)$, and 10 random samples.

	\begin{figure}[t!]
		\centering
		\includegraphics[width=14cm]{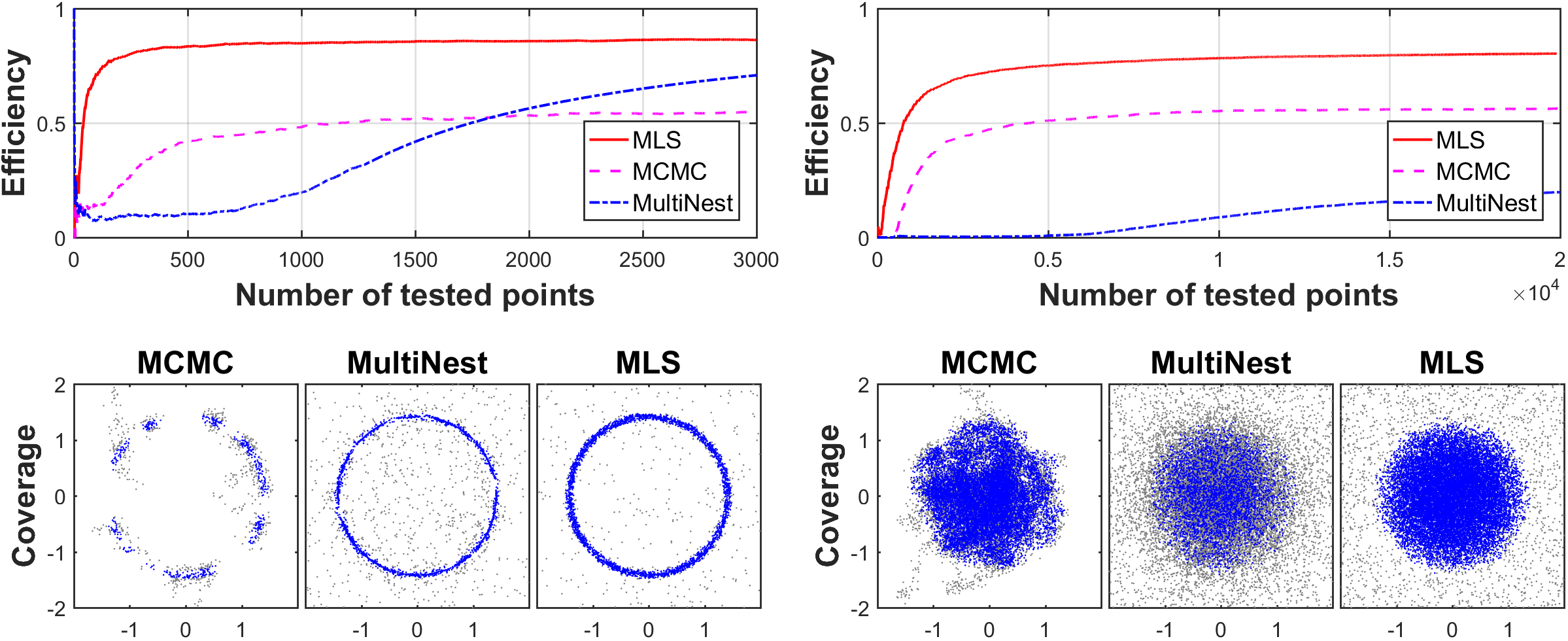}
		\caption{The analysis of the parameter space of the high dimensional toy model:
	       the sample coverage rates and sampling efficiencies of the MLS, the MCMC and the MultiNest
	       scan methods. For the 2-dimensional model (left) and the 7-dimensional model (right)
	       the scan analysis, the scan analysis uses
	       3000 and $2 \times 10^4$ samples, respectively. This figure is taken from \cite{Ren:2017ymm}.}
		\label{Chap3/parabolic-nd}
	\end{figure}

	Fig.~\ref{Chap3/parabolic-nd} shows the sampling efficiencies and the sample distributions of the 2-dimensional and 7-dimensional models given by the MLS, MCMC and MultiNest methods. For the MCMC and MultiNest, the parameters are carefully optimized in order to use as few samples to get the best parameter space coverage. For the MCMC, we use the Gaussian proposed distribution, with the distribution width set to 1/50 of the range of the parameter space, and run 10 independent Markov chains. For the MultiNest, we use 200 and 1000 live points for the 2-dimensional and 7-dimensional models, respectively. From the results we can see that for the same number of sampled parameter points, the sampling efficiency and the sample distribution of the MLS method are the best. In order to compare the performance changes of scan algorithms with the increase of parameter space dimensions, we measured the sampling efficiency of various scan algorithms for different parameter space dimensions with 1000 collected samples. For the MLS method the sampling efficiencies of the 2-dimensional and 7-dimensional models are 84.5\% and 55.5\%, respectively; while for the MCMC they are 48.2\% and 22.9\%, and for the MultiNest they are 19.6\% and 0.4\%, respectively. On the other hand, the distribution of samples obtained by MCMC method is the worst, and it does not cover the survived parameter space to a sufficient extent, because the MCMC chain is easy to be trapped into the local extrema of likelihood function and hard to jump out.

	\subsubsection{MSSM alignment limit}

	Now we apply the MLS method to new physics phenomenology studies. The minimum supersymmetric standard model (MSSM) is currently the most popular extension of the standard model (SM). Under the current LHC constraints and dark matter experiments, when the light CP-even Higgs boson ($h$) acts as the SM Higgs boson, the mass of bino-like dark matter in the MSSM must be above 30~GeV~\cite{Ambrogi:2017lov, Abdughani:2017dqs}. However, in the alignment limit~\cite{Carena:2013ooa}, the heavier CP-even scalar boson ($H$) can also act as the SM Higgs boson with a mass of 125~GeV. Meanwhile, the light CP-even Higgs boson ($h$) can be very light \cite{Bechtle:2016kui, Profumo:2016zxo}. Under the constraint of dark matter relic density, through the interaction of a light $h$, it is possible for a bino-like dark matter below 30~GeV.

	We used MLS and MultiNest to analyze the parameter space of the MSSM to find the alignment parameter space. We require the parameter points to satisfy the constraints:	
	(i) The CP-even $H$ in the MSSM model is taken as the SM-like Higgs boson;
	(ii) Satisfy the experimental values of the Higgs-related physical observables, which is carried out by using HiggsBounds-4.3.1\cite{Bechtle:2013wla} and HiggsSignals-1.4.0~\cite{Bechtle:2013xfa} (yielding the corresponding $\chi^2$ value denoted as $\chi^2_\mathrm{HS}$);
	(iii) Dark matter relic density $\Omega h^2$ should be in the experimental range at $3\sigma$ level~\cite{Ade:2015xua}.

	The masses of the Higgs bosons $m_h$ and $m_H$ are calculated using FeynHiggs 2.13.0 \cite{Heinemeyer:1998yj} and dark matter relic density $\Omega h^2$ using MicrOMEGAs 4.3.2 \cite{Belanger:2010gh}. We performed a parameter space scan in a wide range of parameter space.
	For the first sampling of 200 random parameters, we found that the dark matter relic density $\Omega h^2$ spans several orders (from zero to $10^3$) while other physical observables change rather slowly. Therefore, in our parameter space scan we take the likelihood function as
	\begin{eqnarray}
		\theta(3 - |m_H - 126|) \times \theta(112.7273 - \chi^2_\mathrm{HS}) \times \theta(0.03651 - |\Omega h^2 - 0.1186|) \times e^ {-0.15 m_h}
	\end{eqnarray}
	where the step function $\theta$ means that the physical observable value must strictly fall in the corresponding interval. The last term is used to guide the scan to look for $h$ as light as possible. We used four separate regressors to fit $m_h$, $m_H$, $\chi^2_\mathrm{HS}$ and $\ln\Omega h^2$, respectively, and one classifier for identifying nonphysical parameter points. For both regressors and classifier, fullly connected DNN are used and they have almost the same structure: an input layer, four hidden layers and one output layer. The input layer contains three input neurons corresponding to three free input parameters. Each hidden layer has 60 neurons, and each neuron is activated by the ReLU function. The output neurons of the classifier are activated by the Sigmoid function. The output neurons of the regressors are linear. For training the classifier we choose the binary cross-entropy function as the loss function, while for training the regressors we use the mean-square-error function as the loss function. The standard ADAM optimizer is used to train the learnable parameters of these classifiers and regressors. For each iteration of the MLS algorithm, each classifier and regressor are optimized up to 2000 epochs.

	For the parameter sample recommendation in each iteration, the candidate samples are first filtered by the classifier. Then the selected samples are sent to the regressors, to evaluate their corresponding physical observables and calculate their corresponding likelihoods. If using a physical package, the calculation for a parameter point takes about two seconds, while the DNN can calculate $10^6$ parameter points every second.

	\begin{figure}[t!]
		\centering
		\includegraphics[width=12cm]{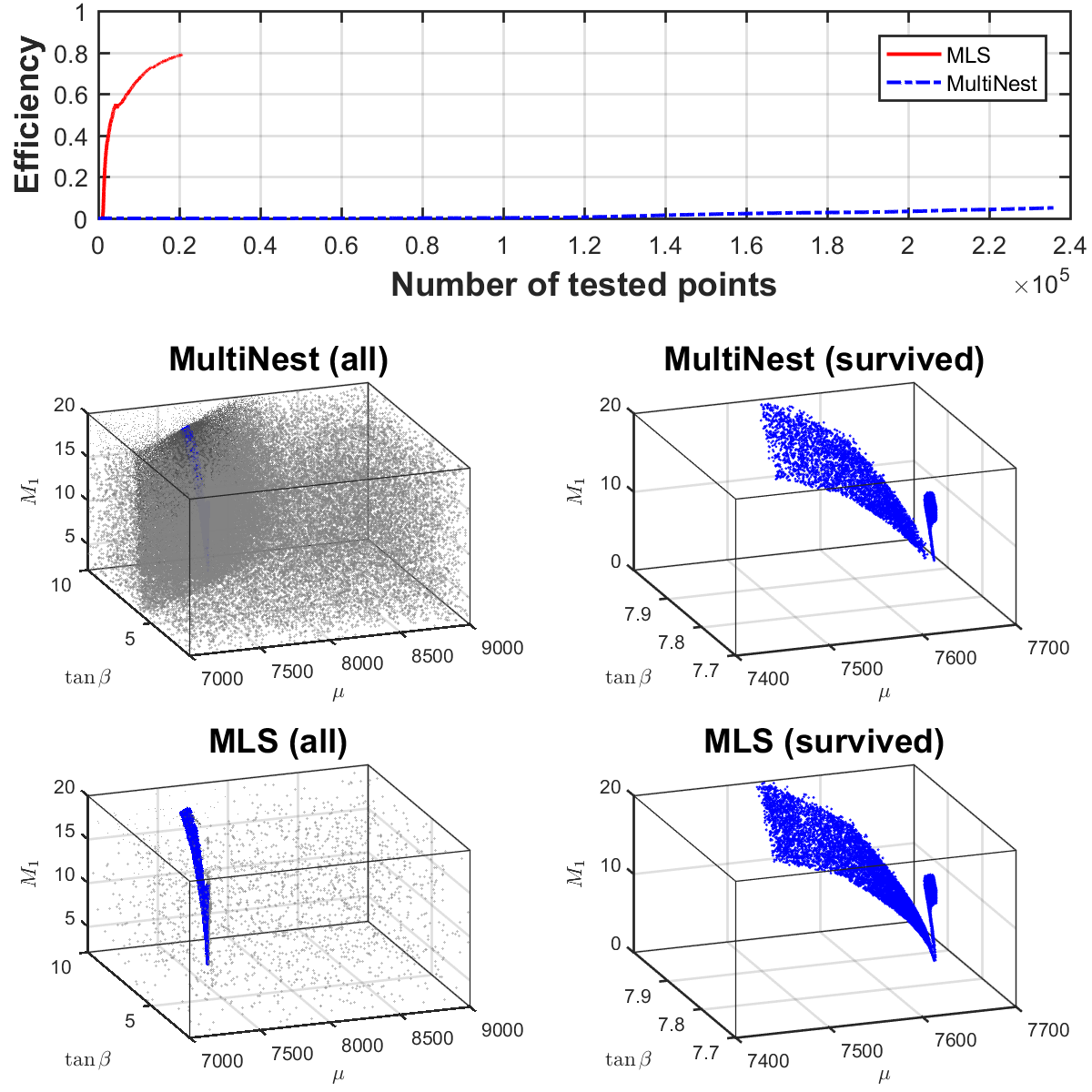}
		\caption{The analysis of the parameter space of the MSSM alignment limit:
	       the sample coverage and sampling efficiencies of the MLS and MultiNest
	       scan methods. This figure is taken from \cite{Ren:2017ymm}.}
		\label{Chap3/mssm}
	\end{figure}

	Our experiments showed that the ML model quickly learns the structures of the physical observables and the likelihood function in the parameter space, so that the subsequent parameter points generated by MLS iterations lie in or around the survived parameter space. As shown in Fig.~\ref{Chap3/mssm}, after 20 iterations, about 90\% of the parameter points generated by each MLS iteration falls within the survived parameter regions. The total MLS sampling efficiency reaches 65\%. The random scan, by contrast, needs to use $10^5$ samples and its sampling efficiency is only 0.68\%. The MLS method is about two orders faster.
	
	In Fig.~\ref{Chap3/mssm}, we show the analysis of the parameter space of the MSSM alignment limit, displaying the sample coverage and sampling efficiencies of the MLS and MultiNest scan methods. There are two independent survived parameter regions in the scan ranges. For the MLS method we use 1000 random parameter points to initialize the DNN. Then, each iteration produces 100 parameter points, 95 of which come from the sampling of the approximate likelihood function, and the other 5 are random parameter points. We see from the results that the MLS sampling efficiency of the parameter points increases very fast, which means that the DNN quickly learns the structure of the physical observable functions and the likelihood function in the parameter space. When sampling $2 \times 10^4$ parameter points, the MLS sampling efficiency has reached 80\%. However, for MultiNest with optimized parameters of algorithm (using 1000 live samples), its sampling efficiency is relatively low, and it needs to keep running till sampling $2.35 \ times 10^5$ parameter points to cover the whole survived parameter regions. Even so, compared with the MLS method, MultiNest scan results have a poor coverage for the survived parameter region with a smaller $M_1$.

	\subsubsection{CMSSM survived parameter space}

Now we use the MLS method to analyze the CMSSM parameter space, and with comparison to the MultiNest results, to verify the effectiveness of our MLS method in the study of actual phenomenological study. Limited by our computing resources, we scan the CMSSM parameter space:
		$ 5~\mathrm{TeV} < M_0 < 10~\mathrm{TeV}$,
		$1~\mathrm{TeV} < M_{1/2} < 10~\mathrm{TeV}$,
		$|A_0| < 10~\mathrm{TeV}$,
		$3 < \tan\beta < 70$ and
		$\mathrm{sign}(\mu) = -1$.
	In addition, all the SM parameters are fixed in our calculation. In order to use the likelihood function in real phenomenological research, we directly use the likelihood function defined in the open source package GAMBIT~\cite{Athron:2017qdc}. The likelihood function contains many contributions from precise electroweak measurements PrecisionBit~\cite{Workgroup:2017bkh}, dark matter detection DarkBit~\cite{Cornell:2017opo}, flavor physics FlavBit~\cite{Workgroup:2017myk} and direct search at colliders ColliderBit~\cite{Balazs:2017moi}. We use the open source package GAMBIT-1.1.3~\cite{Athron:2017qdc} to calculate the corresponding physical observables and the likelihood function of the parameter samples. Among them, the calculations of physical observables use a lot of external packages, such as micrOMEGAs 3.6.9.2~\cite{Belanger:2014vza}, DDCalc 1.0.0~\cite{Workgroup:2017lvb}, FlexibleSUSY 1.5.13~\cite{Athron:2014yba}, gamLike 1.0.0~\cite{Workgroup:2017lvb}, GM2Calc 1.3.0~\cite{Athron:2015rva}, HiggsBounds 4.3.1~\cite{Bechtle:2008jh}, HiggsSignals 1.4~\cite{Bechtle:2013xfa}, SuperIso 3.6~\cite{Mahmoudi:2007vz}, SUSY-HIT 1.5~\cite{Muhlleitner:2003vg} etc.

	\begin{figure}[t!]
		\centering
		\includegraphics[width=10cm]{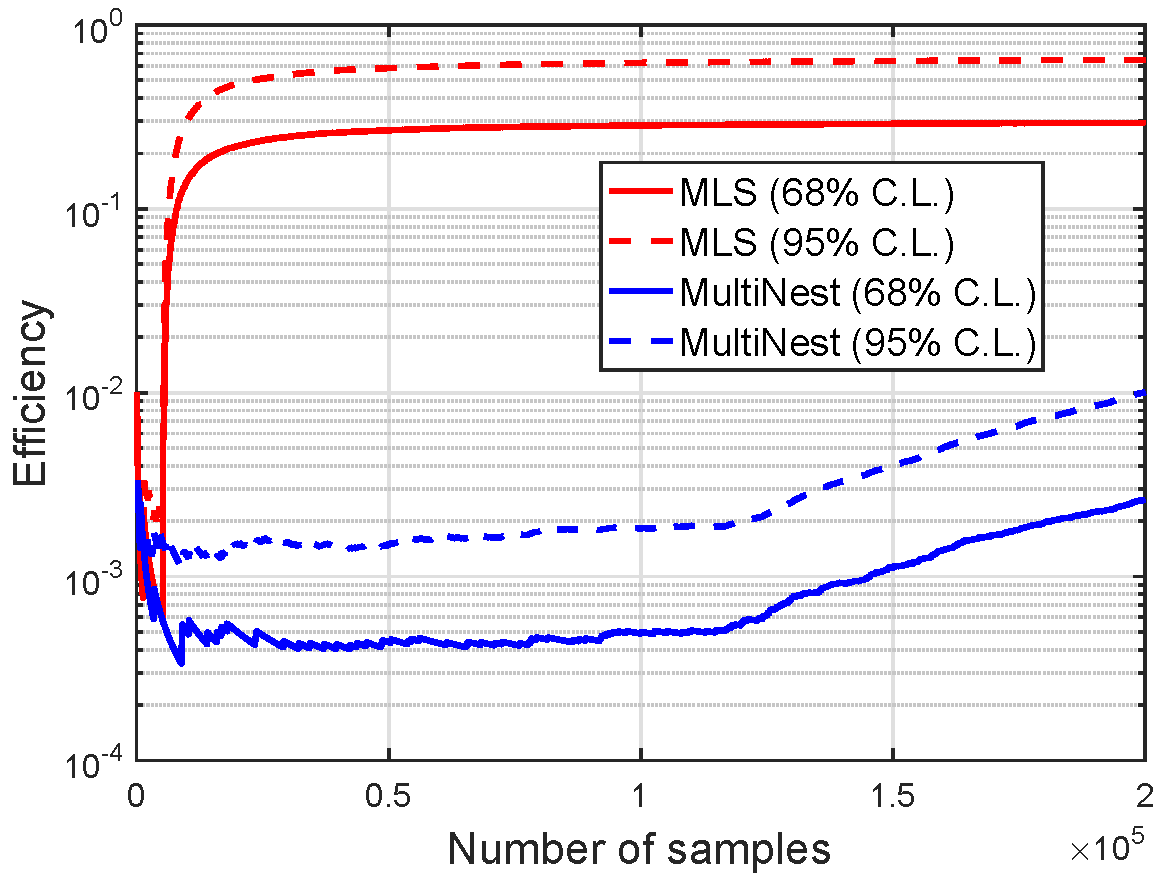}
		\caption{Sampling efficiencies of the CMSSM parameter space: the MLS uses
	                 5000 initial samples and the MultiNest uses 5000 live samples.
	                 Both methods perform the calculation for $2 \times 10^5$ parameter samples. This figure is taken from \cite{Ren:2017ymm}.}
		\label{Chap3/cmssm-effi}
	\end{figure}

	Using the above likelihood function, we initialize the MLS with 5000 random parameter samples. After evaluating 200,000 parameter points, 58,719 (129,249) samples are found to locate within 68\% (95\%) CL region of the parameter space. Actually, the survived regions can be mostly discovered with about 20,000 samples. For comparison, we also performed a MultiNest scan with 5,000 live points, same as the setting in GAMBIT ~\cite{Athron:2017qdc}. It turns out that 523 (2016) out of 200,000 samples are within 68\% (95\%) CL region of the parameter space. Fig.~\ref{Chap3/cmssm-effi} shows the dependence of sampling efficiency on the number of samples. If the number of samples is large enough, the efficiency of MultiNest may become comparable with MLS.

	\begin{figure}[t!]
	\centering
	   \includegraphics[width=14cm]{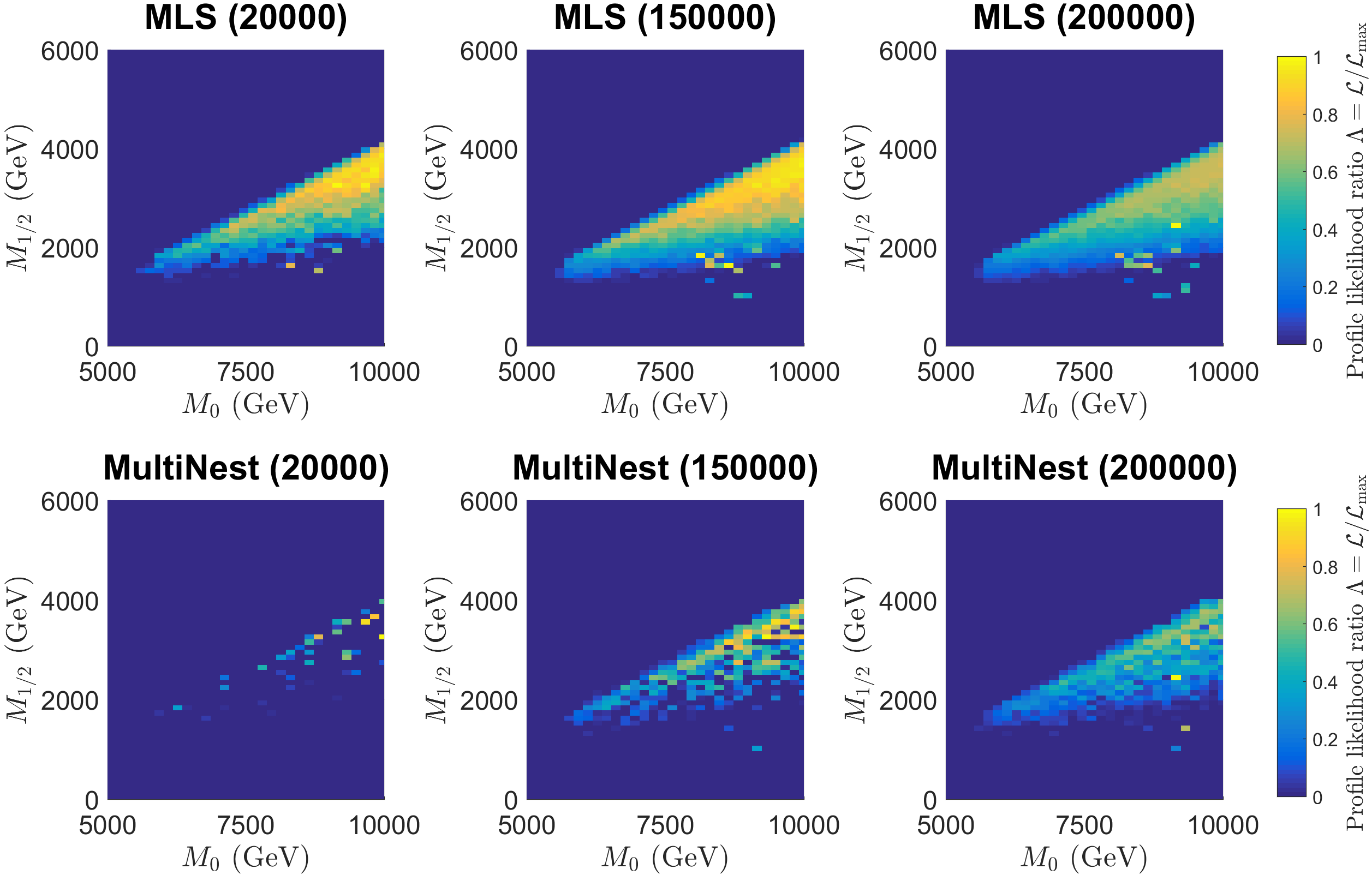}
	   \caption{The profile likelihood ratio displayed on the $M_0 - M_{1/2}$ plane
	            for $2 \times 10^4$, $15 \times 10^4$ and $20 \times 10^4$
	            samples obtained from the MLS (upper panel) and MultiNest (lower panel), respectively. This figure is taken from \cite{Ren:2017ymm}.}
	\label{cmssm_profile}
	\end{figure}

	Fig.~\ref{cmssm_profile} shows the profile likelihood ratio on the $M_0 - M_{1/2}$ plane for $2 \times 10^4$, $15 \times 10^4$ and $20 \times 10^4$ samples obtained from the MLS (top panel) and MultiNest (bottom panel), respectively. In this region, the relic abundance of neutralino dark matter is achieved through the chargino-neutralino coannihilation or the $A/H$ funnel. We see that the chargino-neutralino coannihilation region (the large continuous region in Fig.~\ref{cmssm_profile} for the MLS scan is consistent with the MultiNest scan as well as with the  GAMBIT result~\cite{Athron:2017qdc}. For the $A/H$ funnel region, i.e., the sporadic regions around $M_0\simeq 8~\mathrm{TeV}$ and $M_{1/2}\simeq2~\mathrm{TeV}$, the MLS method can find more samples than the MultiNest. So we successfully show that the MLS has a higher sampling efficiency than MultiNest in case of limited computing resource.

	\begin{figure}[h]
		\includegraphics[width=14.5cm]{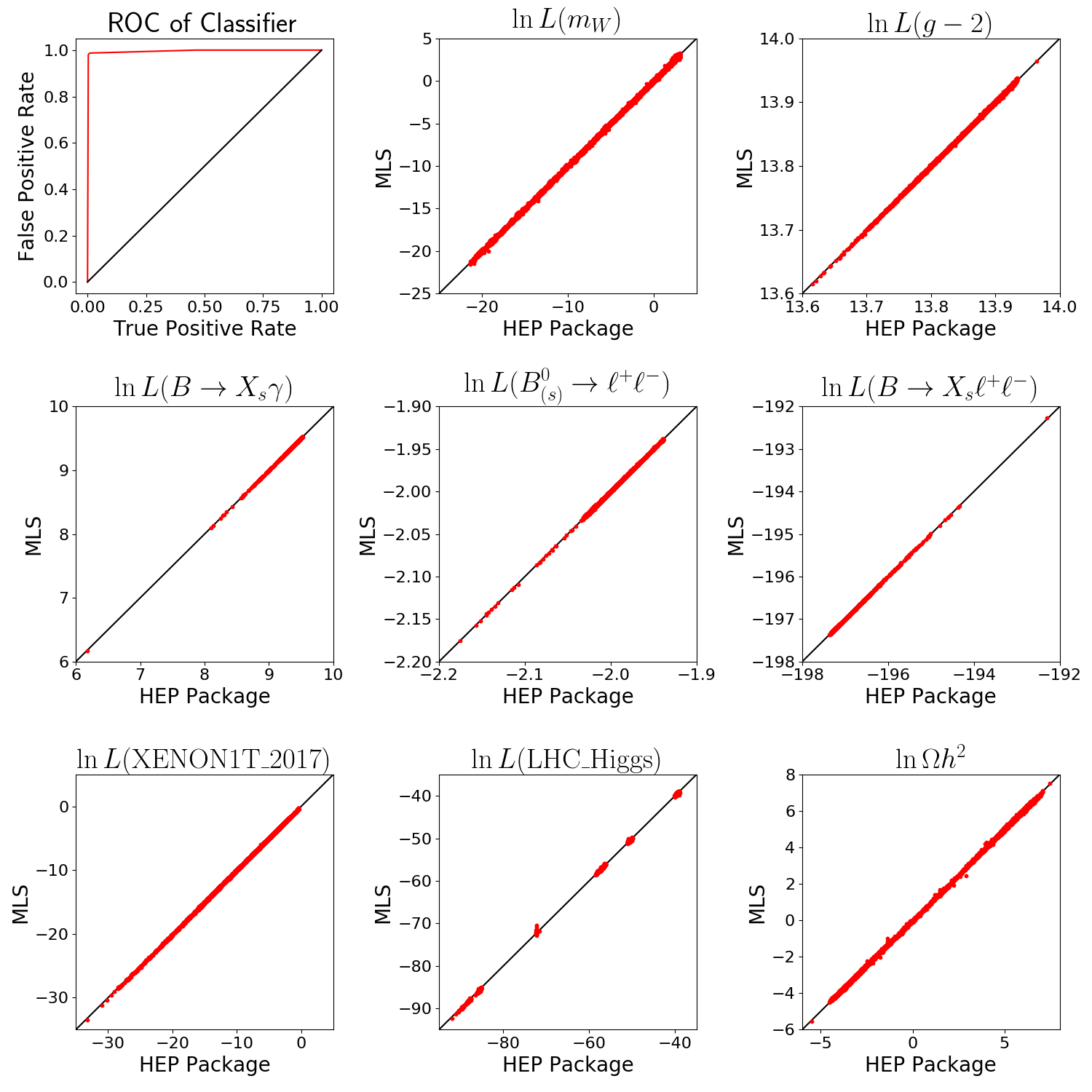}
		\caption{The ROC curve of a deep neural network classifier for
	                 identifying physical and non-physical parameter points (upper left),
	                 and comparion of the experimental physical observable values estimated by the
	                 deep neural network regressors with those calculated by physical packages. This figure is taken from \cite{Ren:2017ymm}.}
		\label{Chap3/cmssm-fit}
	\end{figure}

	In Fig.~\ref{Chap3/cmssm-fit} we show the performance of the ML model in the MLS. We see that the ROC curve of the DNN classifier is very steep, which means that the physical and non-physical parameter points can be distinguished very well. Also we can see that the DNN regressors can accurately estimate the values of the experimental physical observables (or likelihood function) from the parameter points.

	To summarize, we proposed a new method for exploring the parameter space of new physics models, called the machine learning scan (MLS). It can quickly and reliably explore the parameter space of a new physics model with multiple parameters and multiple independent survived parameter regions. As a performance comparison, we compared the MLS method with other conventional scan methods in several toy models and two practical phenomenological studies. It can be seen from the comparison results that the MLS method can greatly reduce the calculation cost required by the parameter space analysis and at the same time improve the sampling efficiency and the coverage of the survived parameter regions.

	\subsection{Graph neural network in new physics studies}

	In addition to the traditional cut-flow based analysis techniques, ML methods provide another way to distinguish signals from background events. So far the traditional ML methods have been used in collider data analysis for about 30 years~\cite{Bhat:2010zz}. An example is that in the LHC experiment the use of BDT in data analysis helps to discover the Higgs particle~\cite{Roe:2004na}. Recently, more ML techniques have been developed and applied to the study of high energy physics~\cite{Baldi:2014kfa, Baldi:2014pta, Bridges:2010de, Buckley:2011kc, Bornhauser:2013aya, Caron:2016hib, Bertone:2016mdy}.

	When using ML technique to deal with collider events, we must first construct a representation of events and then choose a ML model to learn and infer this representation. A collider event is usually described as a collection of final-state particles with specific kinematic features. It is worth noting that the geometric relationship between the final-state particles in the event can be used as a sensitive feature to distinguish the signal from background. Moreover, in mathematics such geometric patterns composed of some entities are usually expressed in the form of graphs, so as to facilitate the further use of ML algorithm for numerical analysis.

	Among various ML methods, the MPNN~\cite{Gilmer2017} provides a universal framework for supervised graph learning. It is especially suitable for problems that need to learn and infer the geometric patterns of graphs. In fact, the MPNN is an extension and improvement of the original Graph Neural Network (GNN)~\cite{Gori05, Scarselli09}. At the same time, the MPNN also improves the training and infering efficiency of the original GNN model. The MPNN can be regarded as a nonlinear mathematical model with learnable parameters which can map the input in graph form directly to the output. We can use supervised learning to train their learnable parameters. At present, some varieties of MPNN have been applied in many fields such as network science, molecular physics and jet physics~\cite{Henrion2017, Gilmer2017}.

	In our work \cite{Abdughani:2018wrw,Ren:2019xhp} we applied the MPNN to the classification of signals and background events at the LHC.
		Each collider event is represented as a graph, which we call the event graph. In an event graph, the nodes describe the final-state particles and the edges represent the geometric relationship between each two final-states in the event.
		Noe that unlike RNN and fully connected DNN, the MPNN is a dynamic neural network which is inherently independent of the number of nodes in the input graph and also independent of the ordering of nodes.
	In the following we provide a detailed description of our applications \cite{Abdughani:2018wrw,Ren:2019xhp} .

	\subsubsection{Event graph neural network}

	{\bf Event graph representation:}
	A collision in a collider usually produces a lot of final-state particles. These final-state particles are further reconstructed into objects such as photons, leptons (electron or muon) and jets with four-momentum information. Each collider event can naturally be represented by a graph, based on the space geometry formed by final-state objects in the detector. We call this representation event graph representation.

	Given a collider event, the corresponding event graph is constructed as follows. Each final object is represented as a node in the graph, and then each pair of nodes is connected by an edge, forming an undirected weighted complete graph. Each node in the event graph has a feature vector. We denote the feature vector of the node $i$ as $\bm{x}_i$, which is used to describe the main information of the final object $i$ in the event. Each edge in the event graph has its own weight. We denote the edge connecting nodes $i$ and $j$ as $d_{ij}$, which is used to represent the geometric distance between two final objects of $i$ and $j$ in the event.

	\begin{figure}[t!]
		\includegraphics[width=14cm]{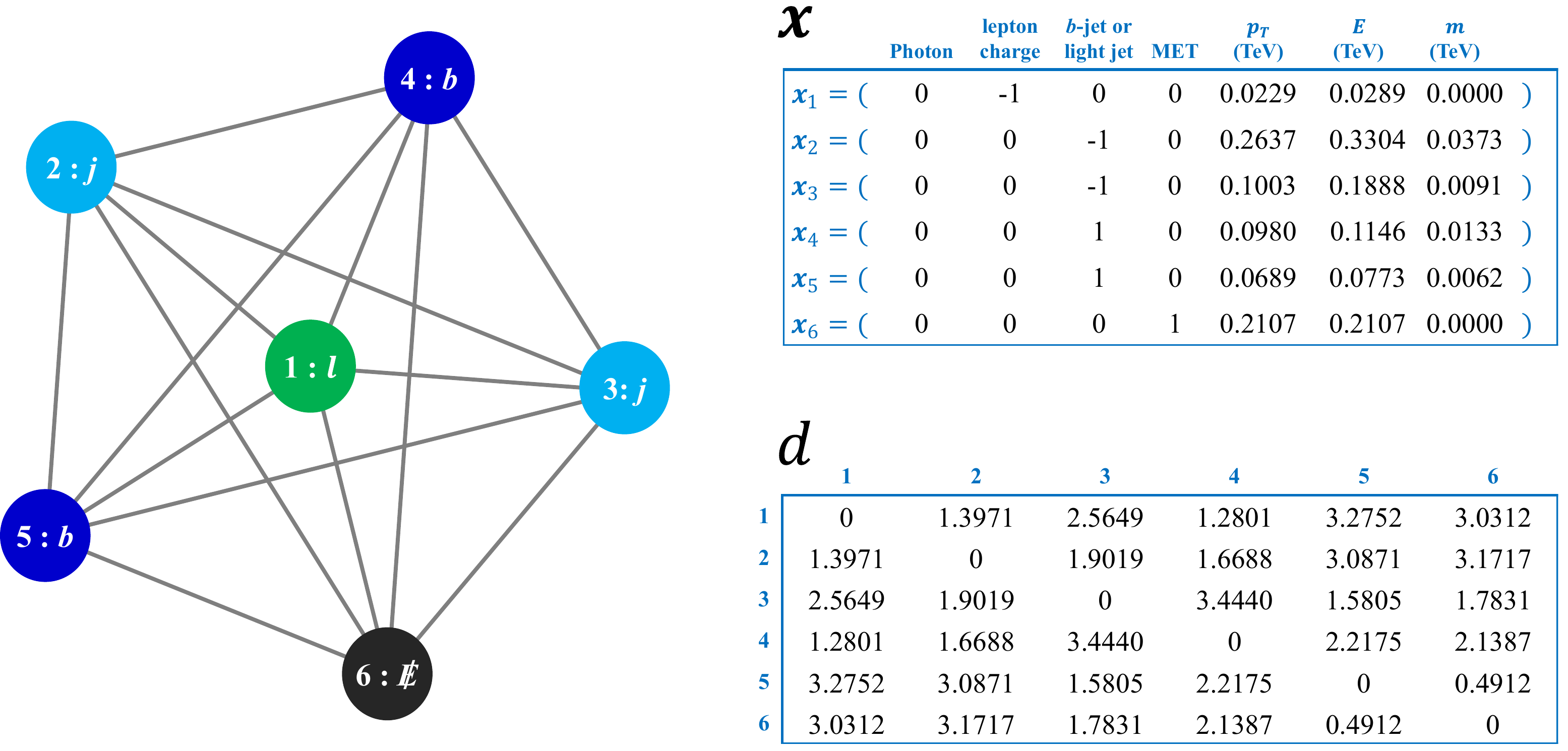}
		\caption{ A sample of event graph from
			the process $pp \to \tilde{t}_1 \tilde{t}^*_1 \to t\bar{t}\tilde{\chi}^0_1\tilde{\chi}^0_1 \to \ell + 2b + \ge 2j + \slashed{E}_T$ at the 13 TeV LHC. This figure is taken from \cite{Abdughani:2018wrw}.}
		\label{event-graph}
	\end{figure}

	As an illustration, in Fig.~\ref{event-graph} we show an event graph of the process $pp \to \tilde{t}_1 \tilde{t}^*_1 \to t\bar{t}\tilde{\chi}^0_1\tilde{\chi}^0_1 \to \ell + 2b + \ge 2j + \slashed{E}_T$, where the detailed node feature vector $\bm{x}$ and edge weights matrix $d$ are also given. To be specific, the main properties of the $i$-th final-state are encoded into the 7-dimensional node feature vector $\bm{x}_i$. The first feature indicates that the final-state is a photon (1) or not (0). If the final-state is a lepton, its charge acts as the second feature; otherwise 0. The third feature indicates that the final-state is a b-jet (1), light jet (-1) or not a jet (0). The fourth feature is used as an indicator of missing energy (MET) (1) or not (0). The rest three features are transverse momentum ($p_T$), energy ($E$) and mass ($m$) of the object. By construction, such a feature vector is simple and compact, and, at the same time, is extendable. By extending the dimension of node  feature vector, various other experimental data associated with final objects can be conveniently added to the feature vector.

	In the event graph, the edge weight of two nodes $i$ and $j$ is set as the angular distance of two final objects $i$ and $j$:
		$d_{ij} = \sqrt{\Delta y_{ij}^2 + \Delta\phi_{ij}^2}$
	with $\Delta y$ and $\Delta \phi$ respectively representing the difference between rapidities and azimuth angles.
	
	It is noteworthy that in the event graph construction we deliberately do not put the information of the angle $\phi$ of the final object in the node feature vector. Instead,  only the azimuth angle difference $\Delta \phi$  between two nodes is included in the edge weight. In this way we can guarantee that the event graph is invariant under the rotation of collider events around the beam axis.

\begin{figure}[t]
	 	\centering
	        \includegraphics[width=14cm]{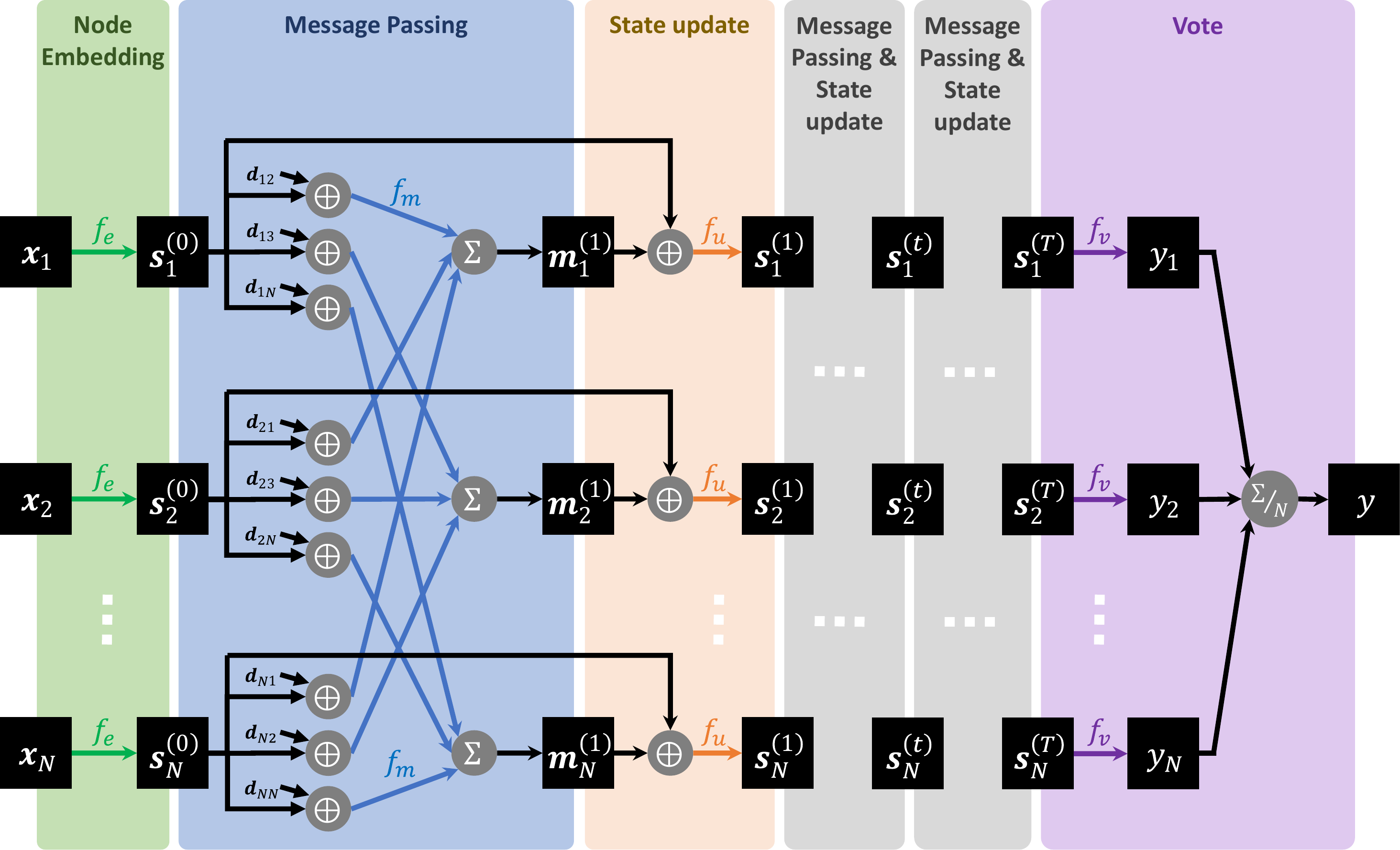}
	        \caption{The structure of our MPNN designed for event graph classification.
	It has the functional layers shown as shadowed blocks,
	with $T$ pairs of message passing and state updating layers for automatic
	event feature extraction. The state vectors $\bm{s}$, the message vectors $\bm{m}$,
	the votes $y_i$ and the discrimination score $y$ are shown as black boxes.
	The colored arrows denote the application of node embedding function $f_e$,
	message passing function $f_m$, state update function $f_u$
	and vote function $f_v$, respectively.
	The operators are given in gray circles, with $\oplus$ denoting
	vector concatenation, $\Sigma$ and $\Sigma/N$ being summation and average, respectively. This figure is taken from \cite{Abdughani:2018wrw}.}
	        \label{mpnn-arch}
	    \end{figure}

	{\bf Graph neural network design:}
	In our work, we designed a variant of MPNN to better classify events. Fig.~\ref{mpnn-arch} shows the basic structure of the MPNN we designed. This NN consists of a series of functional layers, including a node embedding layer, $T$ groups of message passing and node state update layers as well as a voting output layer.

	First, in the node embedding layer the feature vector $\bm{x}_i$ of each node is mapped to a high-dimensional node state vector
		$\bm{s}_i^{(0)} = f_e(\bm{x}_i)$
	with $f_e$ being the node embedding function. Up to now the state vector $\bm{s}_i^{(0)}$ only encodes the $i$-th node features $\bm{x}_i$ without any information about the geometrical pattern of the whole graph.

	Then the message passing techniques are used to perform the event graph embedding, which encodes the whole event graph into each node state vector. This is done by a number of message passing layers and node state update layers. At iteration $t$, each node $i$ collects the messages sent from other nodes $j$
	\begin{equation}
		\bm{m}_i^{(t)} = \sum_{j \neq i} \bm{m}_{i \leftarrow j}^{(t)} = \sum_{j \neq i} f_m^{(t)} (\bm{s}_{j}^{(t-1)}, d_{ij}) ,
	\end{equation}
	and update its state vector
    $\bm{s}_i^{(t)} = f_u^{(t)} (\bm{s}_i^{(t-1)}, \bm{m}_i^{(t)})$
	with $f_m^{(t)}$ being the learnable message functions and $f_u^{(t)}$ being the learnable update functions. By repeating this procedure, the information in node states and the distances between nodes are disseminated with the sent messages, and each node updates its knowledge of other nodes and the relationships between all nodes. Therefore, after $T$ iterations each node state is an encoding of the whole graph, which is a compact representation of the information of both the kinematic features of all final-states and the geometrical relationship between them. They can be viewed as event features automatically extracted from the input event graph.

	Next, based on its own state vector, each node votes a number which is the likeness of the event as the signal
		$y_i = f_v(\bm{s}_i^{(T)})$
	with $f_v$ being the vote function. Finally, to make the prediction stable we take average for
	the votes from all nodes as the final score of discrimination of the event
		$y = \sum_i y_i/|\mathcal{V}|$
	with $|\mathcal{V}|$ being the number of nodes.
    Then the event selection can be made by applying a specific cut $\theta_y$ on the score $y$, i.e., only the events with $y > \theta_y$ are selected.

	In our calculations, the state and message vectors are chosen as 30-dimensional, and the single layer perceptrons are chosen for the node embedding, message passing, update and vote functions
	\begin{eqnarray}
		f_e(\bm{x}) &=& \mathrm{relu} \left( W_e \bm{x} + \bm{b}_e \right) , \\
		f_m^{(t)}(\bm{s}, d) &=& \mathrm{relu} \left( W_m^{(t)} (\bm{s} \oplus [d]) + \bm{b}_m^{(t)} \right) , \\
		f_u^{(t)}(\bm{s}, \bm{m}) &=& \mathrm{relu} \left( W_u^{(t)} (\bm{s} \oplus \bm{m}) + \bm{b}_u^{(t)} \right) , \\
		f_v(\bm{s}) &=& \sigma \left( W_v \bm{s} + \bm{b}_v \right) ,
	\end{eqnarray}
	with $\oplus$ denoting the vector concatenation, $\mathrm{relu}$ being the rectified linear unit, $\sigma$ being the sigmoid function, $W$s and $\bm{b}$s being the learnable parameters.  For each iteration the independent message and update functions are used. To ease the learning of message functions, we expand the pair distance $d$ on a Gaussian basis $N(\mu_i, \delta^2)$, linearly distributed in [0,5] with width of 0.25, as a 21-dimensional vector $[d]_i = \exp\{ (d - \mu_i)^2 / 2 \delta^2 \}$.
	
	We found that the above choices are a good trade-off between model complexity and prediction accuracy. We not only consider to increase the complexity of the neural network model to ensure that it can fully learn and express the characteristics of sample data, but also consider to decrease the complexity of the neural network model to reduce the overfitting phenomenon and improve the generalization prediction ability of the model. Based on the above selection, our MPNN model has a total of 7051 learnable parameters.

	The MPNN can be efficiently trained using supervised learning techniques. We adopt binary-cross-entropy as the loss function and use the Adam~\cite{KingmaB14} optimizer with a learning rate of 0.001 to optimize the model parameters based on the gradients calculated on a mini-batch of 500 training examples. A separate set of validation samples is used to measure the generalization performance. All the calculations are implemented with the open-source deep learning framework PyTorch~\cite{PyTorch} with a strong GPU acceleration.

	\subsubsection{Graph neural network for top-squark search at LHC}

	In order to verify our MPNN method, we apply it to the stop search at the LHC, i.e., the search for the stop pair production $pp \to \tilde{t}_1 \tilde{t}^*_1 \to t\bar{t}\tilde{\chi}^0_1\tilde{\chi}^0_1 \to \ell + 2b + 2j + \slashed E_T$ at the 13 TeV LHC with 36.1~fb$^{-1}$. We assume that the lightest supersymmetric particles $\tilde{\chi}^0_1$ is bino-like, and focus on the kinematic region $m_{\tilde{t}_1} \ge m_t + m_{\tilde{\chi}^0_1}$. The main backgrounds come from $t\bar{t}$, $W+\mathrm{jets}$ and $tW$. The background $t\bar{t}Z(\to \nu \bar{\nu})$  is non-negligible for a heavy stop and is also included in our calculations. The multi-jet background is found to be negligible for this signal \cite{Aaboud:2017aeu}.

	The package \textsf{MadGraph5\_aMC@NLO}~\cite{Alwall:2014hca} is used as an event generator to simulate signal and backgrounds at parton-level. The parton shower and hadronization are carried out with \textsf{Pythia8.2}~\cite{Sjostrand:2014zea}, and a fast detector simulation is performed using \textsf{Delphes-3.4.1}~\cite{deFavereau:2013fsa}. We use the anti-$k_t$ algorithm \cite{Cacciari:2008gp} with the distance parameter $R=0.4$ to cluster jets, and assume a $b$-tagging efficiency of 80\%. The event preselection is performed by \textsf{CheckMATE-2.0.14}~\cite{Drees:2013wra} with the following pre-selection cuts: exactly one lepton with $p_T(\ell) > 10$ GeV and $|\eta(\ell)| < 2.5$, at least 4-jets/2b with $p_T(j) > 25$ GeV and $|\eta| < 2.5$, and the transverse missing energy $\slashed E_T >150$ GeV. The stop pair production rate to the NLO QCD is calculated with \textsf{Prospino}~\cite{Beenakker:1999xh}. The backgrounds $t\bar{t}$ and $W$+jets are normalized with their NNLO cross-sections~\cite{Czakon:2011xx, Boughezal:2015dva}. We display the significance $Z = S / \sqrt{B}$ and, to guarantee the statistics, we require at least 10 events after the cuts.

	In order to compare with the performance of other machine learning methods, we also use deep fully connected neural network (DNN) to classify events. Since the input size of DNN is fixed, we sort the nodes in each event graph by their identities and $p_T$ and then arrange the node features and edge weights into a fixed-size feature vector of the form
	\begin{equation}
		[ \bm{x}_{\slashed{E}_T}^T, \bm{x}_{\ell}^T, \bm{x}_{b_1}^T, \bm{x}_{b_2}^T, \bm{x}_{j_1}^T, \cdots, \bm{x}_{j_N}^T; d_{\slashed{E}_T \slashed{E}_T}, d_{\slashed{E}_T \ell}, \cdots, d_{j_N j_N} ],
		\label{input_for_dnn}
	\end{equation}
	where $N=$ 17 is the maximum number of light jets in our event graphs. Zero-padding is adopted to fill the missing values in the feature vector, namely for an
	event graph with $n$ light jets, $\bm{x}_{j_k} = \bm{0}\ (k > n)$ and $d_{j_k j_l} = 0\ (k\ \mathrm{or}\ l > n)$. We carefully tune the hyperparameters of DNN and training settings to avoid over-fitting. The DNN has 588 input neurons, four 400-neuron hidden layers with the relu activation function and a dropout rate of 0.5, and one output neuron with the sigmoid activation function. It is trained with the same settings as in our MPNN.

 \begin{table}[t!]
        \caption{The significance ($Z$) of MPNN and DNN for two benchmark points at 13 TeV LHC with the luminosity of ${\cal L}=36.1$ fb$^{-1}$. MPNN6 and DNN6 are for the results of using six objects
        	(one lepton, two b-jets, two leading light-jets and MET) as inputs.}
        \label{comparison}
        \centering
        \begin{tabular}{|c|c|c|c|c|c|c|}
            \hline\hline
    & ~$m_{\tilde{t}_1}$ (GeV)~ &  $~m_{\tilde{\chi}^0_1}$ (GeV)~&  ~$Z$(MPNN)~ & ~$Z$(DNN)~& ~$Z$(MPNN6)~ &~$Z$(DNN6)~
\\ \hline
  ~point A~ & 525 & 352& 4.6 & 2.7 & 3.5 & 2.9 \\ \hline
  ~point B~ & 900 & 330& 5.4 & 4.0 & 4.3 & 4.2 \\
            \hline\hline
        \end{tabular}
    \end{table}

	Table~\ref{comparison} shows the performances of MPNN and DNN for two benchmark points. The two benchmark points are chosen with distinctive kinematic features: point A is in the compressed region $m_{\tilde{t}_1} \approx m_t + m_{\tilde{\chi}^0_1}$, while point B is in the uncompressed region $m_{\tilde{t}_1} \gg m_t + m_{\tilde{\chi}^0_1}$.  The results of MPNN6 and DNN6 with six objects (one lepton, two b-jets, two leading light-jets and MET) as inputs are also presented.
	 The reduced event graphs for MPNN6 have 6 nodes and 36 edges. The input feature vectors for DNN6 are of the form of Eq.~(\ref{input_for_dnn}) with $N=2$ so that it has 78 input features.
	From Table~\ref{comparison} we can see that the MPNN has a better significance than DNN for both benchmark points, especially for the case $A$ in which the significance increases from $2.9\sigma$ (DNN6) to $3.5\sigma$ (MPNN6), and $2.7\sigma$ (DNN) to $4.6\sigma$ (MPNN). With more input features, MPNN outperforms over MPNN6, while DNN has no such a feature, since more learnable parameters usually lead to a more serious over-fitting for DNN.

	\begin{figure}[t!]
		\centering
		\includegraphics[width=14cm]{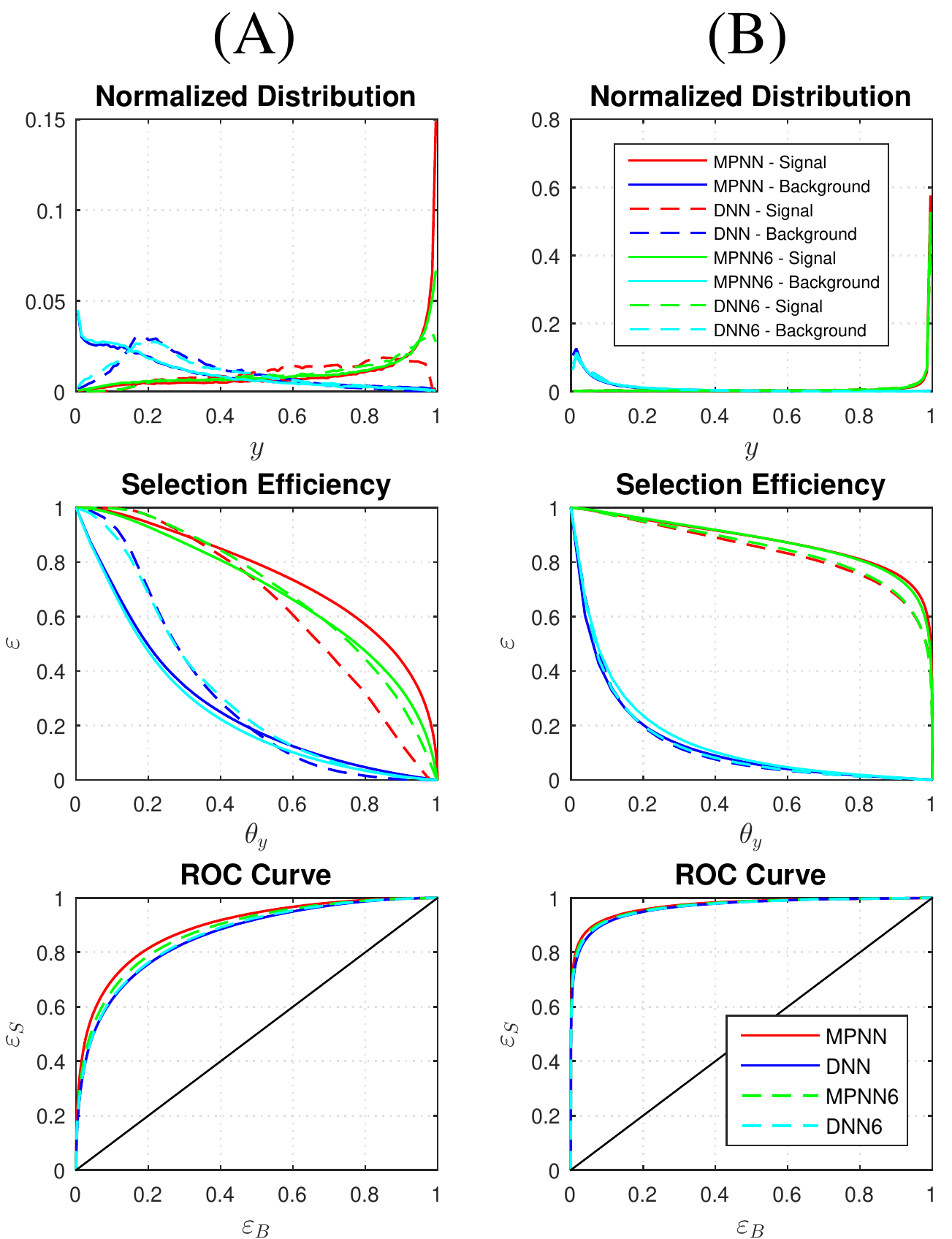}
		\caption{The discriminating powers of MPNN and DNN on signal and backgrounds for
		benchmark points A and B in Table~\ref{comparison}. This figure is taken from \cite{Abdughani:2018wrw}.}
		\label{benchmark}
	\end{figure}

	\begin{figure}[t!]
		\centering
		\includegraphics[width=10cm]{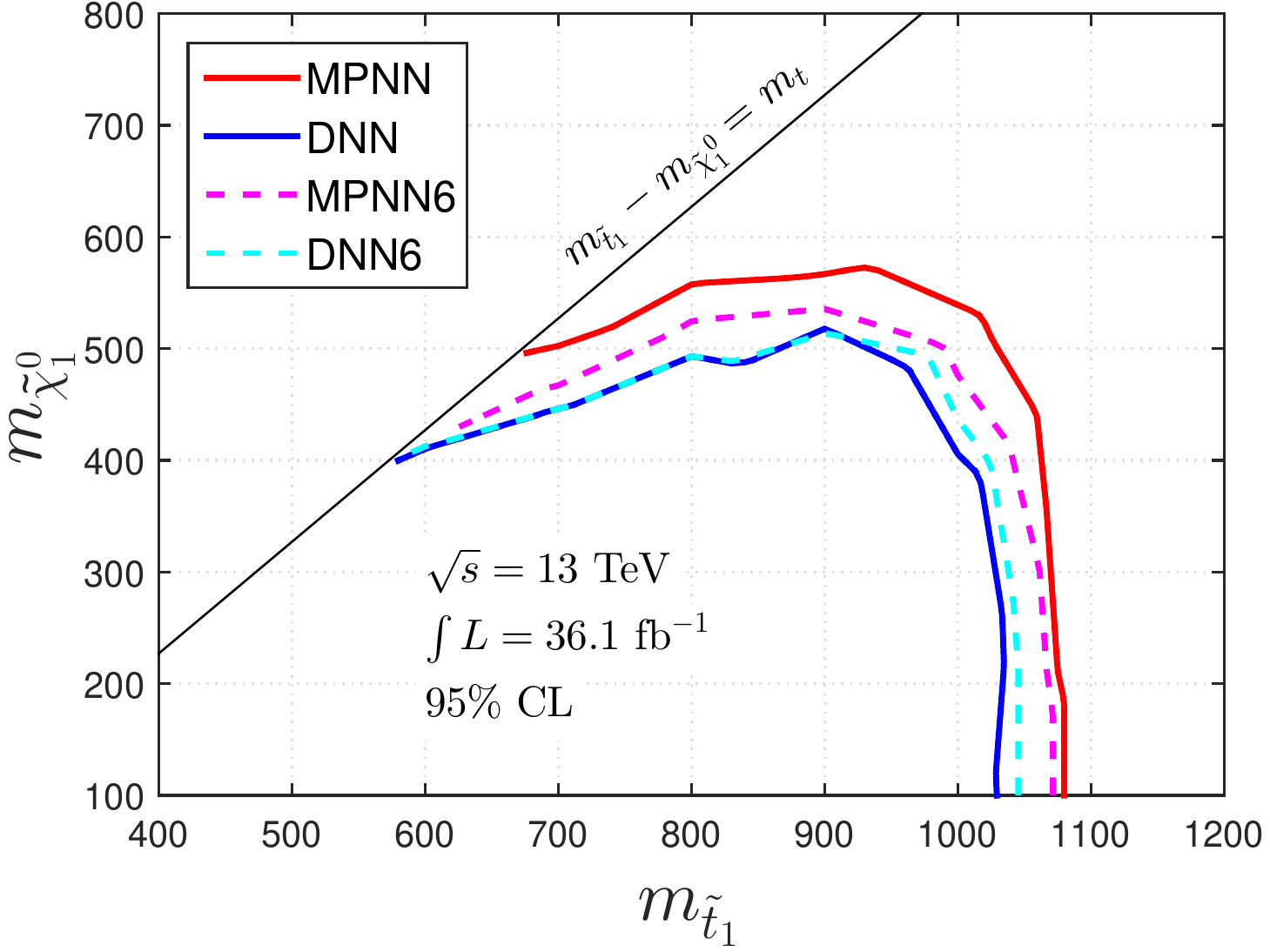}
		\caption{The 95\% CL exclusion limits on the plane of $m_{\tilde{t}_1}$ versus $m_{\tilde{\chi}^0_1}$
			by using MPNN and DNN at the 13 TeV LHC with a luminosity of ${\cal L}=36.1$ fb$^{-1}$.
			The black line corresponds to the mass relation $m_{\tilde{t}_1}=m_{\tilde{\chi}^0_1}+m_{t}$. This figure is taken from \cite{Abdughani:2018wrw}.}
		\label{exclusion}
	\end{figure}

	Fig.~\ref{benchmark} shows the discriminating power of MPNN and DNN on signal and background events for benchmark points A and B. Since the over-fitting is very small, we only show the results on validation set. The top panel shows that the signal and background events are well separated in the distributions of discrimination scores for both MPNN and DNN. But for MPNN the scores of signals are inclined to have larger values than DNN, while the scores of backgrounds have smaller values than DNN. The middle panel shows that MPNN has a higher signal selection efficiency $\varepsilon_S$ and lower background selection efficiency $\varepsilon_B$ than DNN. This leads to sharper receiver operating characteristic (ROC) curves than DNN, as shown in the bottom panel.

	Fig.~\ref{exclusion} shows the 95\% C.L. exclusion limits on the plane of $m_{\tilde{t}_1}$ versus $m_{\tilde{\chi}^0_1}$ by using MPNNs and DNNs at the 13 TeV LHC with the luminosity ${\cal L}=36.1$ fb$^{-1}$. We see that MPNNs can produce stronger exclusion limits than DNNs. For examples, in the compressed region $m_{\tilde{t}_1} \approx m_{\tilde{\chi}^0_1} + m_{t}$, the mass bound of stop from MPNN can be 670 GeV, which is about 100 GeV larger than the DNN result. In other regions  $m_{\tilde{t}_1} > m_{\tilde{\chi}^0_1} + m_{t}$, the exclusion limits of stop mass can be enhanced by several tens of GeV.

	\subsubsection{Graph neural network for probing top-Higgs coupling at the LHC}
	
	An accurate measurement of the coupling of the top quark and the Higgs boson will allow for a deep understanding of new physics beyond SM. In our work \cite{Ren:2019xhp}, we used the MPNN to study the CP property of this coupling through the semi-leptonic decay channel of the process $pp \to t \bar{t} H$ at the 13~TeV LHC. Based on the event classification probability of the output of the MPNN, we constructed a variable and performed a hypothetical test. We found that the pure CP-even and CP-odd couplings can be well distinguished at the LHC with at most 300~fb$^{-1}$ experimental data.

	Including the possible CP-violation, the top-Higgs interaction can be parameterized as
		 $-(y_t/\sqrt{2}) \bar{t}(\cos\xi + i \gamma_5 \sin\xi) t H$
	with $y_t = \sqrt{2} m_t / v$ and $\xi = 0$ in the SM~\cite{AguilarSaavedra:2009mx} ($v = 174$~GeV is the vacuum expectation value of the Higgs field).  The CP violation in this coupling is caused by the presence of $\sin\xi$ term, which has been constrained~\cite{Cirigliano:2016njn,Kobakhidze:2016mfx}. These indirect constraints are model-dependent and the most robust test of this coupling is from the direct measurement of $t\bar{t}H$ production at colliders~\cite{Gunion:1996xu, Ellis:2013yxa, Bramante:2014gda, Demartin:2014fia, Aguilar-Saavedra:2014kpa, Godbole:2015bda, Buckley:2015vsa, Li:2015kaa, Li:2017dyz, Cao:2018}.
	In the following, we denote the Higgs boson with the CP-even coupling ($\xi = 0$) as $h$, and the one with the CP-odd coupling ($\xi = \pi/2$) as $A$. At the LHC the $t\bar{t}h$ and $t\bar{t}A$ productions have the same backgrounds, dominantly from the process $pp \to t\bar{t}b\bar{b}+X$.

\begin{figure}[t]
		\centering
		\includegraphics[width=14cm,height=8cm]{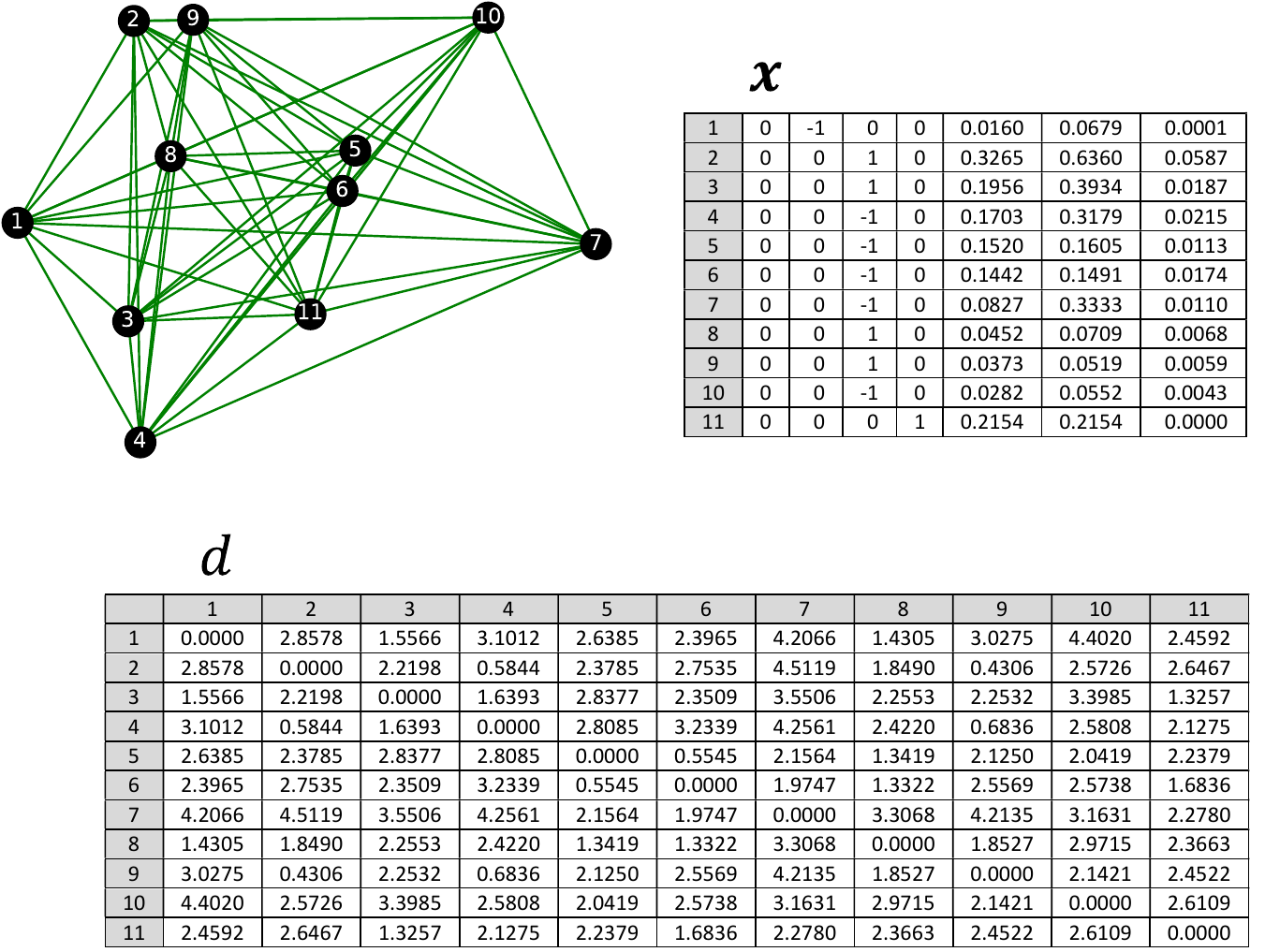}
		\caption{An event graph for a simulated $t\bar{t}h$ event at the 13 TeV LHC. This figure is taken from \cite{Ren:2019xhp}.}
		\label{event-graph-htt}
	\end{figure}

	In our study we used event graphs as the representation of collider events and designed an MPNN to classify the collider events, i.e., give the probabilities of the input event graph $e$ as $t\bar{t}h$, $t\bar{t}A$ or $t\bar{t}b\bar{b}$ denoted as $p(h|e)$, $p(A|e)$ and $p(b|e)$, respectively. Finally, a variable from the output of MPNN is constructed and a hypothetical test is performed. As the input of MPNN, each collider event is represented as an event graph as shown in FIG.~\ref{event-graph-htt}.

	If the top-Higgs coupling is CP-even (CP-odd), the event sample collected in experiments will come from the $t\bar{t}h$ ($t\bar{t}A$) process plus the dominate $t\bar{t}b\bar{b}$ background process.
	To discriminate event samples of the two scenarios, we construct two likelihoods from the single-event probability output from the MPNN
	\begin{eqnarray}
		L_h(D) = {\prod_{e \in D}}' p(h|e),~~~~~~
		L_A(D) = {\prod_{e \in D}}' p(A|e)
	\end{eqnarray}
	 For the CP-even (CP-ood) scenario, we have $L_h(D) \gg L_A(D)$ ($L_h(D) \ll L_A(D)$).
	Then we use a log-likelihood ratio 	$\ln Q(D) = \ln( L_A(D)/L_h(D))$
	to perform a hypothesis test as the test statistics
	The distribution of $\ln Q$ denoted as $f_h$ and $f_A$ in the two scenarios can be numerically obtained by evaluating a large number of random simulated event samples.

	\begin{figure}[t]
          \centering
		\includegraphics[width=12cm]{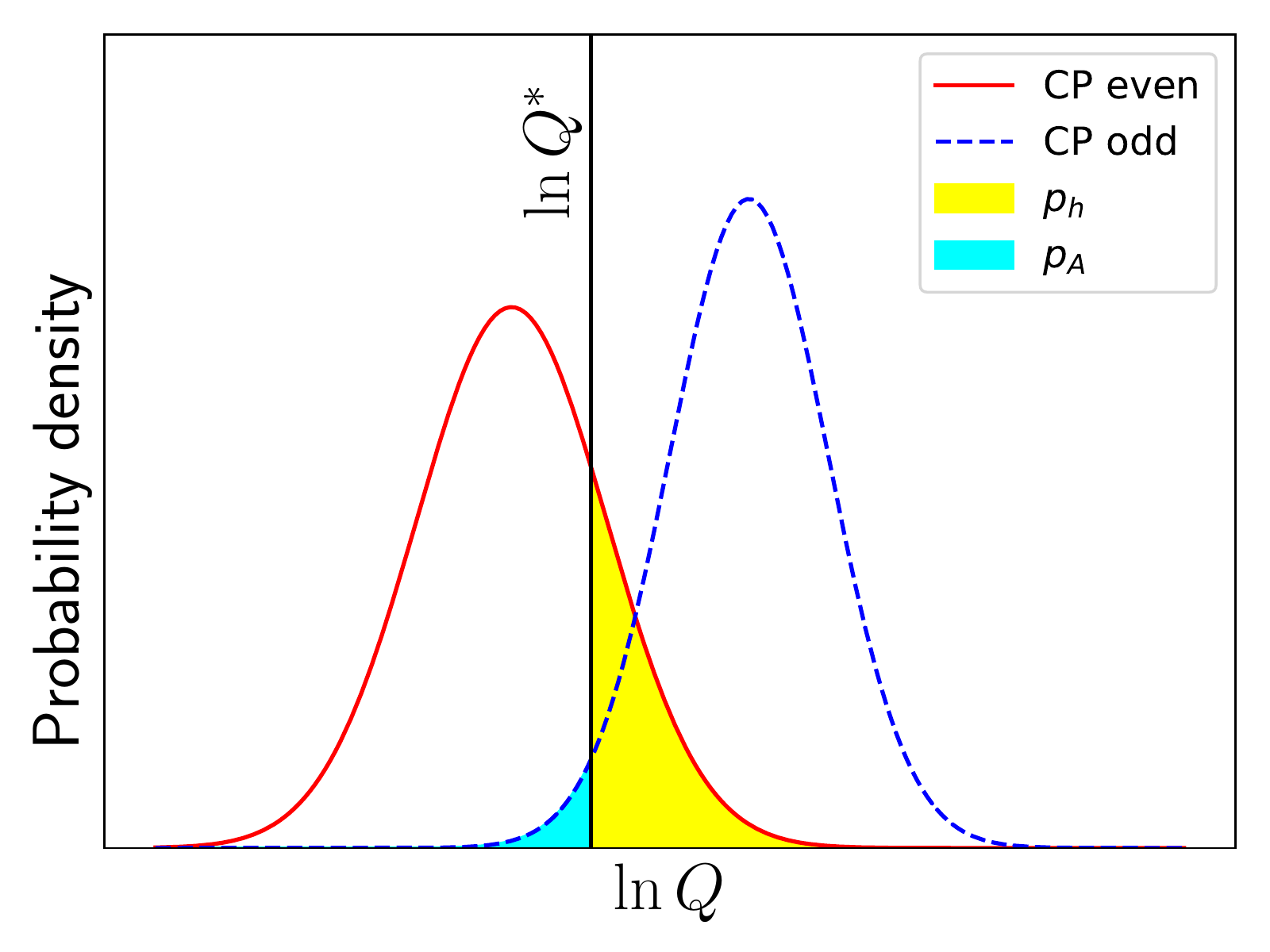}
		\caption{An illustration of $p$-value evaluation for rejecting the CP-even or CP-odd scenario. This figure is taken from \cite{Ren:2019xhp}.}
		\label{p-value-calc}
	\end{figure}

	Once the distributions of $\ln Q$ are given and the value of $\ln Q^*$ calculated from the observed experimental data $D^*$, as illustrated in FIG.~\ref{p-value-calc}, we can evaluate the $p$-values of rejecting the CP-even or CP-odd scenario by the integrals
	\begin{eqnarray}
		p_h(\ln Q^*) = \int_{\ln Q^*}^{+\infty} f_h(x) dx, ~~~~
		p_A(\ln Q^*) = \int_{-\infty}^{\ln Q^*} f_A(x) dx
	\end{eqnarray}

	\begin{figure}[t]
		\centering
		\includegraphics[width=12cm]{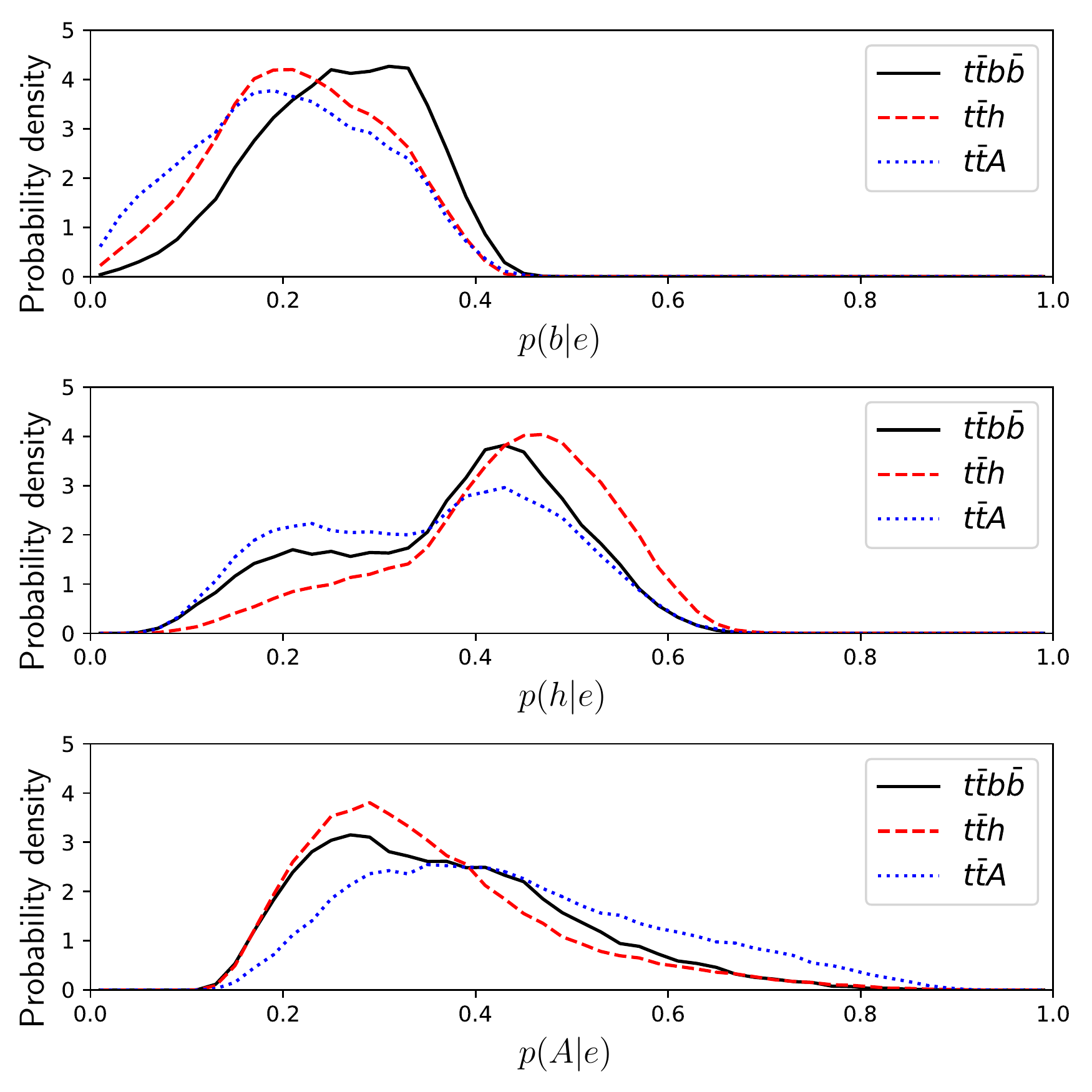}
		\caption{The distributions of $p(h|e)$, $p(A|e)$ and $p(b|e)$ for $t\bar{t}h$, $t\bar{t}A$ and $t\bar{t}b\bar{b}$ events from the MPNN output. This figure is taken from \cite{Ren:2019xhp}.}
		\label{p-dist}
	\end{figure}

	As in the preceding section, in our simulations we used the packages MadGraph5~\cite{Alwall:2014hca},  Pythia8~\cite{Sjostrand:2014zea},
	Delphes~\cite{deFavereau:2013fsa} and CheckMATE2~\cite{Drees:2013wra}.
	We require leptons to have $p_T > 20$~GeV and $|\eta| < 2.5$,
	jets to have $p_T > 25$~GeV and $|\eta| < 2.5$, and assume a b-tagging
	efficiency of  60\%. We used the semi-leptonic channel, which has exactly one lepton, four $b$-jets and at least two light jets in the final states.
    We collected $9\times 10^5$ samples with balanced numbers of $t\bar{t}h$, $t\bar{t}A$ and $t\bar{t}b\bar{b}$ events as the training sets, while another $3\times 10^5$ samples
    as the validation set for the performance evaluation.

	Fig.~\ref{p-dist} shows the distributions of the trained MPNN  outputs evaluated on the validation set. We see that the MPNN has successfully learned some discriminative event features for different processes. The background events tend to have higher $p(b|e)$, while the $t\bar{t}h$ and $t\bar{t}A$ events have higher $p(h|e)$ and $p(A|e)$, respectively.

	\begin{figure}[t]
		\centering
		\includegraphics[width=14cm]{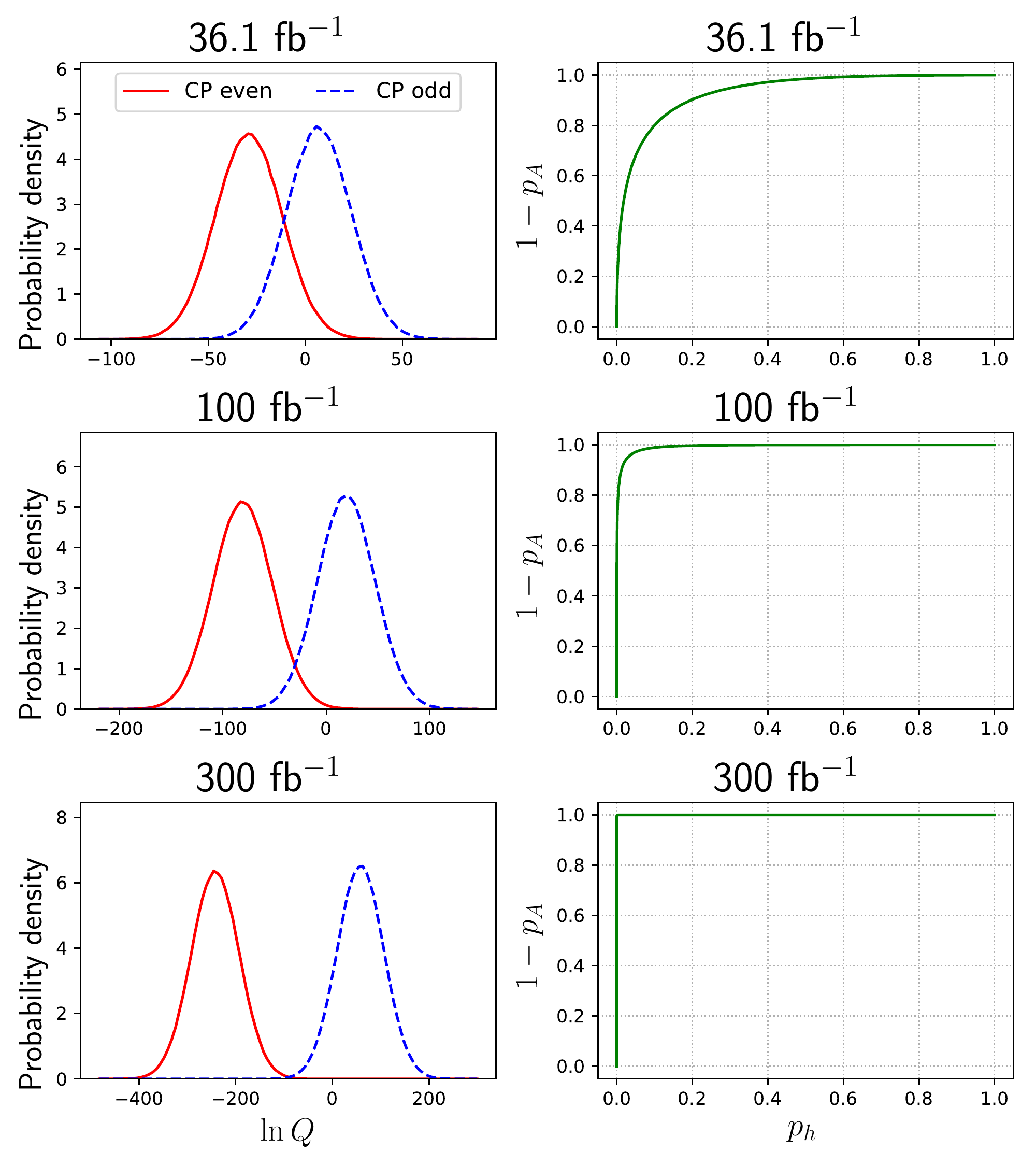}
		\caption{The left panel shows the distribution of the log-likelihood ratio from pseudo-experiments while the right panel shows the receiver operating characteristic (ROC) curve versus $p_h$ of the hypothesis test. This figure is taken from \cite{Ren:2019xhp}.}
		\label{q-dist-roc}
	\end{figure}

	 For each of the two scenarios, millions of pseudo-experiments were performed, with the
	 pseudo-experiment events taken from the validation set. As shown in Fig.~\ref{q-dist-roc},
	 with the increase of luminosity, the distributions for CP-even and CP-odd couplings get more separated, and when the luminosity reaches 300 fb$^{-1}$, the two distributions almost have no overlap.  This indicates that the LHC with 300 fb$^{-1}$ experimental data may distinguish
     the CP property of the top-Higgs couplings using such a MPNN method.
	
	\section{Summary }
	
	Machine learning has been developed as a typical interdisciplinary field. It studies algorithms and statistical models, in order to gradually improve the capabilities of computer systems to solve given tasks. As a sub-field, deep learning techniques have shown powerful learning den inference capabilities which are comparable to or even superior than human experts. Traditional machine learning techniques have already been widely applied to high energy experiments for more than three decades with fruitful results. A great successful application is the using of boosted decision trees in data analysis that leads to the discovery of Higgs boson at the LHC. Currently, deep learning attracts more attentions in high energy experiments, but also some interesting applications has been done using deep learning in the field of phenomenology. In this note we provided a brief review on these applications. We first described various deep neural network models and then recapitulated their applications to high energy phenomenological studies. Some detailed applications including the machine learning scan in the exploration of parameter space, the graph neural networks in the search of top-squark and $Ht\bar{t}$ productions at the LHC were delineated.

	\section*{Acknowledgments}
	
	We thank Rong-Gen Cai for suggestions and comments.
	This work was supported by the National Natural Science Foundation of China (NNSFC) under
	grant Nos. 11705093, 11305049, 11675242, 11821505 and 11851303,
	by Peng-Huan-Wu Theoretical Physics Innovation Center (11747601),
	by the CAS Center for Excellence in Particle Physics (CCEPP),
	by the CAS Key Research Program of Frontier Sciences
	and by a Key R\&D Program of Ministry of Science and technique under number 2017YFA0402200-04.


%

\end{document}